\documentclass[aps,prb,twocolumn,citeautoscript,superscriptaddress]{revtex4-1}  
\pdfoutput=1      
\usepackage{amsmath,amssymb,mathrsfs,bm,feynmf,setspace,xspace,soul}   
\usepackage{graphicx}  
\usepackage[tight]{subfigure}       
\usepackage{color}   
\usepackage[colorlinks=true]{hyperref}            
\hypersetup{
    bookmarks=true,         
    unicode=false,          
    pdftoolbar=true,        
    pdfmenubar=true,        
    pdffitwindow=false,     
    pdfstartview={FitH},    
    pdftitle={My title},    
    pdfauthor={Author},     
    pdfsubject={Subject},   
    pdfcreator={Creator},   
    pdfproducer={Producer}, 
    pdfkeywords={keyword1} {key2} {key3}, 
    pdfnewwindow=true,      
    colorlinks=true,       
    linkcolor=magenta, 
    citecolor=blue,        
    filecolor=magenta,      
    urlcolor=cyan           
} 
 
\newcommand{\de}{\delta}

\newcommand{\ka}{\kappa}

\newcommand{\ba}{\begin{array}{ccc}}
\newcommand{\ea}{\end{array}}
\newcommand{\nn}{\nonumber \\}

\newcommand{\bml}{\begin{multline}}

\newcommand{\eeqm}{\end{multline}}

\newcommand{\bsp}{\begin{split}}
\newcommand{\esp}{\end{split}}

\renewcommand{\b}[1]{{\bf #1}}

\newcommand{\inv}{^{-1}}

\newcommand{\ts}{\thinspace{}}
\newcommand{\req}[1]{Eq.\thinspace(\ref{#1})} 

\newcommand{\rfig}[1]{Fig.\ts\ref{fig:#1}}

\newcommand{\ie}{{\em i.e.\/}\@\xspace} 
\newcommand{\eg}{{\em e.g.\/}\@\xspace}

\DeclareMathOperator{\chord}{ch}

\newcommand{\bl}[1]{#1}     

\newcommand{\thA}{\theta} 
\newcommand{\thc}{\vartheta} 
\newcommand{\asp}{b} 
\newcommand{\tor}{\chi}
\newcommand{\cyl}{\chi^{\rm cyl}}
\newcommand{\torEMI}{\chi_{\rm \scriptscriptstyle EMI}} 
\newcommand{\tortEMI}{\chi^{3d}_{\rm \scriptscriptstyle EMI}} 
\newcommand{\tort}{\chi^{3d}}
\newcommand{\kappat}{\kappa^{3d}}
\newcommand{\infcyl}{\gamma}  
\newcommand{\tEMIsp}{\widetilde{\tor}_{\rm \scriptscriptstyle EMI}}
\begin{document}  

\title{Cornering gapless quantum states via their torus entanglement}
 \author{William Witczak-Krempa} 
 \affiliation{Department of Physics, Harvard University, Cambridge, MA 02138, USA}
\author{Lauren E. Hayward Sierens}
\author{Roger G. Melko}
 \affiliation{Department of Physics and Astronomy, University of Waterloo, Ontario, N2L 3G1, Canada}
 \affiliation{Perimeter Institute for Theoretical Physics, Waterloo, Ontario N2L 2Y5, Canada}
 \date{\today}
\begin{abstract}     
The entanglement entropy (EE) has emerged as an important window into the 
structure of complex quantum states of matter. We analyze the universal part of the EE for gapless 
systems put on tori in 2d/3d, denoted by $\tor$. Focusing on scale invariant systems,  
we derive general non-perturbative properties for
the shape dependence of $\chi$, 
and reveal surprising relations to the EE associated with corners in the entangling surface.
We obtain closed-form expressions for $\chi$ in 2d/3d within a model that arises in the study
of conformal field theories (CFTs), and use them to obtain ansatzes without fitting parameters for
the 2d/3d free boson CFTs. Our numerical lattice calculations show that the ansatzes are highly accurate.
Finally, we discuss how the torus EE \bl{can act as a fingerprint of exotic states such as}
gapless quantum spin liquids, e.g.\ Kitaev's honeycomb model. 
\end{abstract}  
\maketitle   
Measures of quantum entanglement have emerged as powerful tools to characterize complex
many-body systems\cite{Calabrese1,Casini1,Fradkin_book,Wen_book2,laflorencie}, 
such as phases with topological order, gapless  
spin liquids and       
quantum critical states lacking long-lived excitations. The entanglement entropy (EE)
and its R\'enyi relatives have proven especially useful. The EE of a spatial region $A$, heuristically, measures
the amount of entanglement between the inside of $A$ and the outside. Different regions will reveal
different properties about the physical state. Generally, a convenient choice is to work on a space that is periodic
in at least one direction, \ie a cylinder or, particularly in the case of finite-size lattice calculations, a torus.
In this setting, region $A$ is often chosen to wrap around at least  
one cycle, making it topologically non-trivial. 
In a large class of topologically ordered systems in 2 spatial dimensions (2d), the EE of the groundstate
on a cylinder or torus reveals a wealth of information\cite{Dong08,Zhang12,Cincio} about the fractionalized excitations (anyons).   
Furthermore, these EEs have proved to be useful diagnostics in the search for such exotic 
phases\cite{isakov,Wang11,jiang12,depenbrock}.     
In contrast, for gapless states,       
analytical\cite{Fradkin_book,Fradkin06,Stephan09,Hsu09,Max09,Hsu10,Oshikawa10,Max11,Swingle12,fradkin,Pretko} and numerical\cite{Stephan09,Ju_2012,Inglis_2013,Bohdan,Helmes2014,Luitz,Laflorencie2015} studies have revealed      
that the situation is more intricate and numerous open questions remain.                

\begin{figure}
\center 
   \includegraphics[scale=.37]{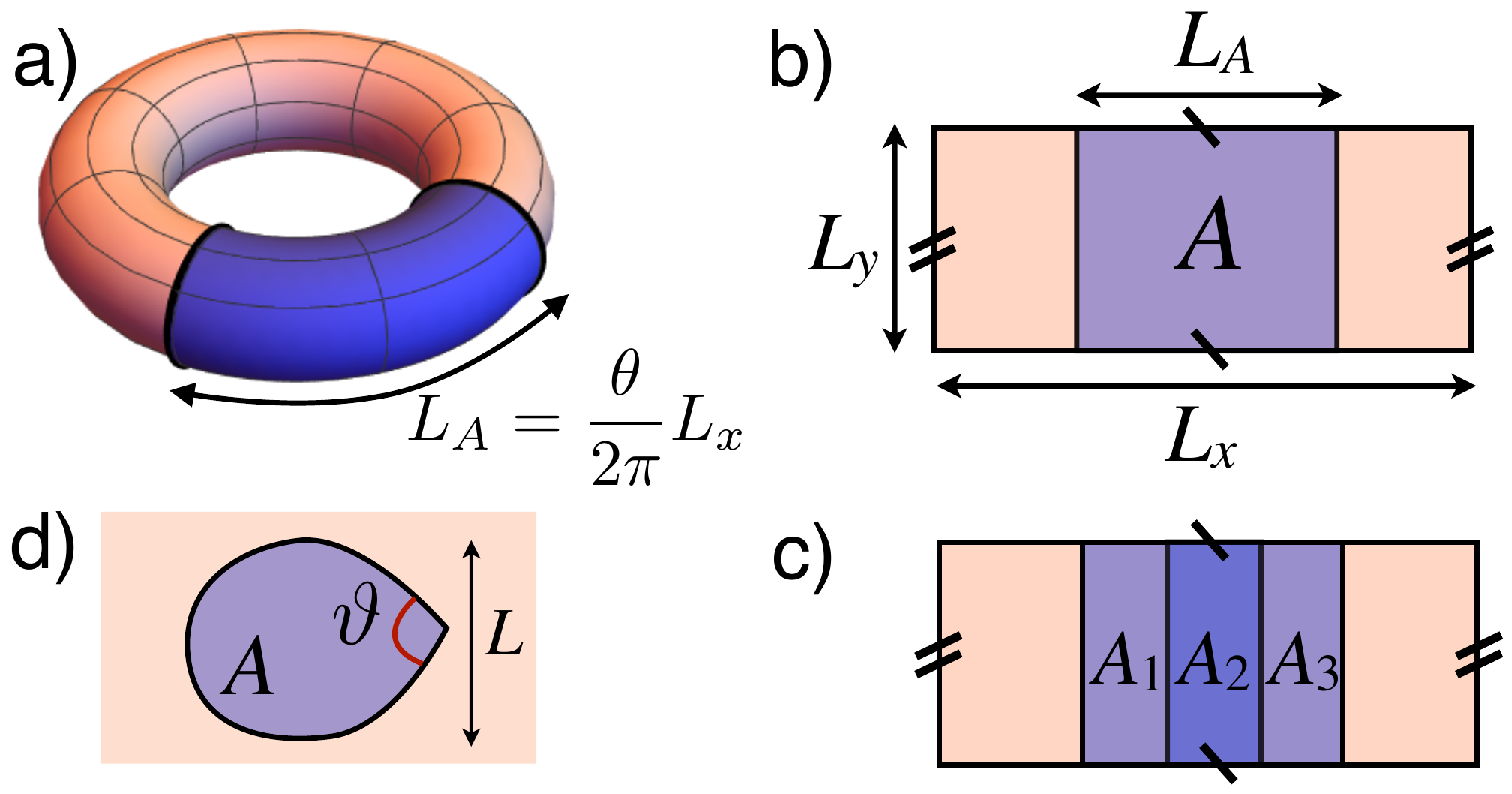} 
   \caption{{\bf a \& b)} 2d space with a torus topology. We study the EE
of a cylindrical region $A$. {\bf c)} Constraints on the torus EE function $\tor(\thA)$ 
result from dividing $A$ into 3 parts, and applying strong subadditivity. {\bf d)} Region with a sharp
corner.}  
\label{fig:torus}
\centering         
\end{figure}

\begin{figure}
\center
   \includegraphics[scale=.99]{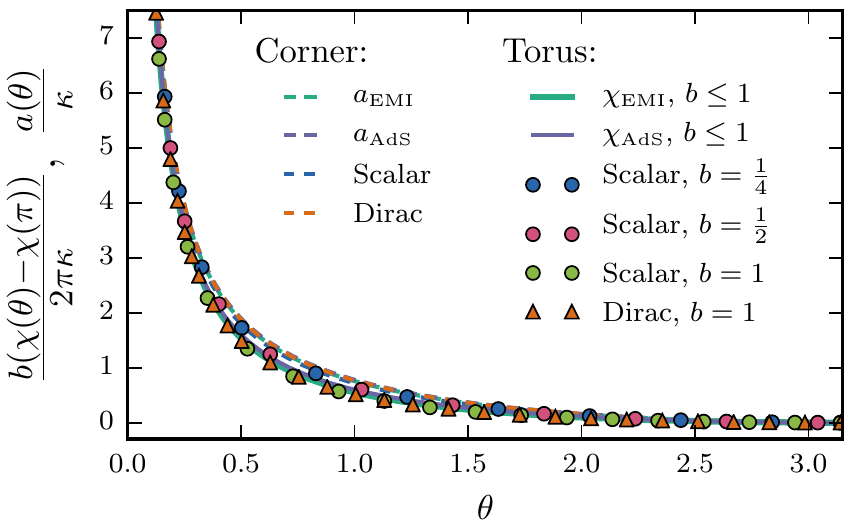}
   \caption{Comparing the universal torus and corner EE of various CFTs in 2d.} 
\label{fig:comparison}  
\centering         
\end{figure}

In this work, we analyze the universal torus EE of gapless theories in 2d/3d.  
We focus our attention on scale invariant systems such as conformal field theories (CFTs)
and Lifshitz quantum critical theories ($z\neq 1$), thus excluding the extra complexity due to Fermi surfaces.
We derive general properties of the torus EE in 2d/3d using strong subadditivity\cite{Lieb73} and other considerations. 
We then make new connections between the shape dependence of the torus EE, and the EE associated
with sharp corners\cite{Casini1}, as shown in \rfig{torus}d. The comparison is natural because both quantities are expressed in terms 
of an angle variable. Surprisingly, we find that the angle dependence of both the torus and corner functions 
are nearly equal when properly normalized, \rfig{comparison}. 
This is illustrated using free CFTs, and strongly coupled ones. 
To gain more intuition about the shape dependence of universal term, we derive a closed-form expression for the torus EE in 2d/3d using a CFT construction.
This allows us to make approximate predictions for the free boson CFT, \emph{without any fitting parameters}.
Our numerical analysis shows that these predictions work accurately.  
We then discuss how the torus EE can be used to reveal both the topological 
and geometrical degrees of freedom of gapless spin liquids, using the Kitaev model as an example. 

{\bf Fundamentals of torus entanglement:}
We consider a system on a flat torus, \rfig{torus}, \ie we identify 
the coordinate $r_i$ with $r_i\!+\! L_i$, $i=x,y$. Given the corresponding groundstate, 
we study its EE associated with 
a cylindrical region $A$ of length $L_A$, $S(A)=-{\rm tr}(\rho_A\ln \rho_A)$;
$\rho_A$ is the reduced density matrix of $A$. The EE scales as 
\begin{align} \label{SA}
  S(A)= \mathcal B\, 2L_y/\de -\tor +O(\de/L_y),
\end{align}
in the limit where $L_i,L_A$ far exceed the microscopic (UV) 
scale $\delta$, which can be taken to be the lattice spacing.
The first term corresponds to the ``area law'',
with a non-universal prefactor $\mathcal B$.  
Our interest lies in the $\de$-independent
term, $-\chi$, because it is \emph{universal}. It remains constant with growing $L_y$, at fixed  
ratios $L_A/L_i$, but in general depends non-trivially on both ratios. $\chi$ thus constitutes a non-trivial measure of the low-energy degrees of freedom of the system, and as we shall see, acts as fingerprint of the state.

We now obtain non-perturbative properties of the torus function $\tor(\thA;\asp)$,  
where we have defined the \bl{natural} angular variable $\thA\!=\!2\pi L_A/L_x$, and the aspect ratio 
$\asp\!=\!L_x/L_y$ (we shall often keep the $b$-dependence implicit).
First, since we are dealing with pure states, the EE of $A$ must equal
that of its complement, \ie $\chi(\thA)=\chi(2\pi- \thA)$; we shall henceforth restrict ourselves to $0\!<\!\thA \!\leq\!\pi$, as in \rfig{comparison}.  
Further, since the limit  where $A$ approaches half the torus is not singular,  
$\tor$ will be analytic about $\pi$: 
\begin{align} \label{exp}
  \tor(\thA\approx\pi) =  \sum_{\ell=0} c_\ell \cdot(\pi-\thA)^{2\ell},
\end{align}  
where only even powers appear due to the aforementioned reflection symmetry about $\pi$. The $c_\ell$ 
depend on the aspect ratio $\asp$, and it would be interesting to understand
which properties of the state they encode. 
To derive further constraints on $\tor$, we invoke an important property of the EE, namely its strong subadditivity\cite{Lieb73} (SSA), 
which implies the following inequality for 3 non-overlapping regions: 
$S(A_1\!\cup\! A_2\!\cup\! A_3)+S(A_2)\leq S(A_1\!\cup\! A_2)+S(A_2\cup A_3)$.  
The key idea is to divide $A$ into 3 regions as in \rfig{torus}c, with angles $\thA_i$, 
and apply SSA.  
Substituting \req{SA} into the SSA inequality, we find that the boundary law contributions cancel and we
are left with $\tor(\thA_1+\thA_2 +\thA_3)+\tor(\thA_2)\geq\tor(\thA_1\!+\!\thA_2)+\tor(\thA_2\!+\!\thA_3)$. 
From this, we can derive
\begin{align} \label{ineq} 
  \tor'(\thA)\leq 0 \,, \qquad \tor''(\thA) \geq 0,
\end{align}
for $0\!<\!\thA\!\leq\!\pi$, \ie the torus function $\tor(\thA)$ is
convex decreasing on that interval. 
As a direct consequence of the inequalities (\ref{ineq}), the second expansion coefficient
in \req{exp} satisfies $c_1\geq 0$ (for all aspect ratios). 

\bl{We now examine the limit $\thA\!\to\!0$ with $L_{x,y}$ fixed, in which case 
the EE reduces to that of a (periodic) thin strip of width $L_A\!\to\!0$
and length $L_y\!\gg\! L_A$. 
We argue that the periodicity in the $x,y$-directions and the associated boundary conditions
do not influence $\tor$ in this limit since the EE is dominated by degrees of freedom  
that do not exceed length scales $\sim\! L_A\ll L_{x,y}$. The total $\chi$ can be obtained by adding 
the contributions from these local patches, and will be proportional to $L_y/L_A$.
We can thus relate the thin slice limit on the torus  
to the EE of a thin strip in \emph{infinite} space. 
For scale invariant systems, this reads\cite{Casini1}
$S_{\rm strip}=\mathcal B 2L/\de- \kappa L/L_A$, 
where $L_A$  is the strip's width. 
$L$ is the long-distance regulator of the infinite strip; 
alternatively, we can define the EE per unit length, $S_{\rm strip}/L$.
$\tor$ will thus have \bl{the same $\kappa L_y/L_A$} divergence in the thin slice limit:}    
\begin{align}  \label{thin}
  \tor(\thA\to 0) = \kappa\,\frac{L_y}{L_A} = \frac{2\pi\kappa}{b\,\thA}.
\end{align}
\bl{Further, by virtue of (\ref{ineq}), $\kappa\geq 0$.}
This means that in the small-$\theta$
limit the full EE, \req{SA}, decreases since the universal contribution $\chi$  
appears with a negative sign. This is consistent since when $A$ vanishes, $S\!=\!0$. 
\bl{The universal constant $\kappa$ has been computed for certain critical theories\cite{Casini_rev};
it will play a central role in our discussion.}

{\bf Relation to corner entanglement:} The above properties share striking similarities with 
the EE associated with sharp corners, as we now explain. 
Given a region $A$ \bl{in the infinite plane} that contains a corner with opening  
angle $\thc$, \rfig{torus}d, the EE scales as
\begin{align}
  S(A)= B\,L/\de - a(\thc) \ln(L/\de) + \dotsb, 
\end{align}
where $B$ is the area law prefactor, and $a(\thc)$ is a \emph{universal} coefficient arising from the  
corner\cite{Casini3,Fradkin06,Hirata07,Casini_rev,Kallin13,Kallin14}. It encodes   
rich low-energy information about the   
state\cite{Fradkin06,corner-prl,corner-twist,Kallin13,Gustainis,Helmes2014,Helmes2,free-corners,laflorencie},    
but in contrast to $\chi$, it vanishes for gapped systems and is thus blind to purely topological degrees of freedom.              
$a(\thc)$ is also symmetric about $\pi$ (at which     
point the corner disappears), and can be expanded as in \req{exp}. 
For CFTs, the leading term in the expansion  
is\cite{corner-prl,Bueno2,faulkner15} $(\pi^2C_T/24)(\pi-\thc)^2$, where $C_T$ 
determines the 2-point function of the stress tensor (and thus of the energy density) in the groundstate. 
\rfig{comparison} shows $a(\thc)$ for the free scalar/Dirac fermion\cite{Casini_rev} and
holographic CFTs\cite{Hirata07}.
Further, $a(\thc)$ obeys the \bl{same} monotonicity and convexity conditions\cite{Hirata07} (\ref{ineq}). 
Finally, in the sharp corner limit $\thc\to 0$, the corner function shows a $1/\thc$ 
divergence\cite{Casini_rev} just as $\chi$: 
$a(\thc\!\to\! 0)\!=\! \kappa_c/\thc$. For CFTs, $\kappa_c\!=\!\kappa$ is exactly the same universal 
constant that controls the divergence 
of $\chi(\thA\!\to\! 0)$, \req{thin}. This holds because the sharp corner geometry can be conformally
mapped to that of a thin strip\cite{Bueno2}, which controls $\tor(\thA\!\to\! 0)$ as discussed above.  
\bl{It would be interesting to see if non-conformal critical theories ($z\!\neq\! 1$) have the same relation between 
their sharp-corner $\kappa_c$ and thin-slice coefficients $\kappa$.}

\begin{figure} 
\center
   \includegraphics[scale=.99]{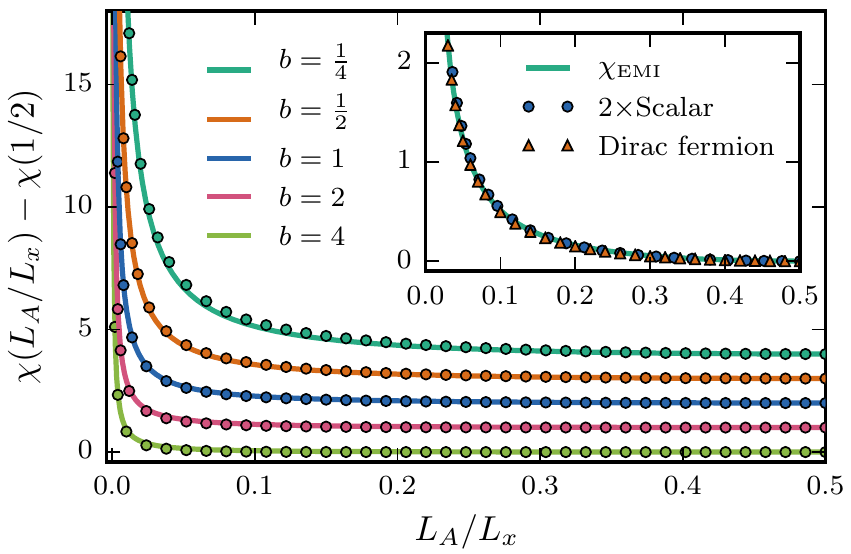} 
   \caption{{\bf Main}: Torus function $\tor$ for the massless scalar in 2d for various aspect 
ratios $\asp=L_x/L_y$; it has been vertically offset 
for clarity. 
The points are numerical data, and the lines are the predictions obtained using $\torEMI$, \emph{without any fitting parameters}.
{\bf Inset}: Dirac fermion data\cite{fradkin} at $b\!=\!1$, the corresponding $\torEMI$, and the complex scalar data
for comparison. The axes represent the same quantities as in the main plot.}
\label{fig:scalar_2d}  
\centering         
\end{figure}

Given the similar asymptotics of $\tor(\thA)$ and $a(\thc)$, one can wonder how they compare at intermediate angles. 
\rfig{comparison} shows the torus and corner functions of various CFTs.
For a meaningful comparison, we normalize them by the thin-slice/sharp-corner coefficient $\kappa$.
Surprisingly, all curves nearly overlap in the entire range of angles. What makes the collapse more 
remarkable is that the curves for the holographic CFTs\cite{fradkin} and the Extensive Mutual Information 
model\cite{Casini05,Casini08} (defined below)    
hold for \emph{all} aspect ratios $b\leq 1$ \bl{(App.~\ref{ap:properties_EMI})}, a non-trivial fact in itself.   
The same $b$-independence of $b\tor$ approximately holds for the massless scalar, as we illustrate with numerical data  
at $b\!=\! 1,\tfrac{1}{2},\tfrac{1}{4}$, taken from \rfig{scalar_2d}. The reason for the collapse constitutes
an open question beyond the scope of this work, \bl{but it suggests a deeper relation between wavefunctions 
on spaces with different topologies/geometries.} 

{\bf Ansatz from extensive mutual information:}
To gain further intuition about the EE on tori, we derive a closed-form ansatz for 
$\tor(\thA)$ that can be meaningfully compared with a large class of gapless states, particularly CFTs.  
To do so we use the Extensive Mutual Information model (EMI)\cite{Casini05,Casini08,Swingle10}, which has proven useful in the analysis
of the EE of CFTs in various dimensions\cite{Casini05,Casini08,Swingle10,corner-prl,corner-twist}. 
The EMI is not defined through a Hamiltonian, but instead allows for a simple geometric computation of the EE within  
the bounds of conformal symmetry, and has passed numerous non-trivial tests\cite{Casini05,Swingle10,corner-prl,corner-twist}.  
The resulting EE of the EMI 
can be interpreted\cite{Swingle10} in terms of an ansatz for twist (or swap) operators used to  
compute R\'enyi and entanglement entropies. 
The designation EMI comes from the fact that its mutual information $I(A,B)=S(A)+S(B)-S(A\cup B)$  
is extensive: $I(A,B\cup C)=I(A,B)+I(A,C)$. 
In infinite flat space, the EE of a region $A$ can be computed as follows within the EMI: 
\begin{align}  \label{emi} 
  S(A) = \int_{\partial A}\! d \b r_1 \int_{\partial A} \! d\b r_2 \,\,\hat n_1\cdot \hat n_2 \, C(\b r_1-\b r_2), 
\end{align}
where $\hat{\b n}$ denotes the unit normal to the boundary $\partial A$, and 
$C(\b r)=s_1/|\b r|^{2(d-1)}$. 
The coordinates $\b r_{1,2}$ live on $\partial A$, and  
$s_1$ is a positive constant. 
In order to apply the prescription (\ref{emi}) to the torus, we need to account for 
the periodicity when determining the function $C$. 
\bl{However, contrary to the infinite plane, conformal invariance and the extensivity of the mutual information
do not suffice to fix $C$ on the torus, and one is left with a richer set of possibilities.
A simple choice for $C$ is described in App.\ts\ref{ap:EMI}; the resulting torus EE reads:}  
\begin{align}  \label{tor-emi} 
  \torEMI(\thA)\! = \!4\kappa\! \left[ \frac{\cot\inv\!\left(\tfrac{\asp}{\pi}\thA\right)}{\asp\,\thA} \! + \!
  \frac{\cot\inv\!\left(\tfrac{\asp}{\pi}(2\pi-\thA)\right)}{\asp\,(2\pi-\thA)} \right]\!
+ 2\infcyl\! 
\end{align}
where $\cot\inv z$ is the inverse cotangent, and $\infcyl$ is a constant.
We have normalized the first term of (\ref{tor-emi}) using $\kappa$ so as to reproduce the 
expected small $\thA$ divergence, \req{thin}.  
$\torEMI$ is thus non-negative for all angles and aspect ratios. 
Our result is naturally symmetric and analytic about $\thA=\pi$, as in \req{exp}, and obeys the constraints (\ref{ineq})
from SSA. \req{tor-emi} thus provides a closed-form candidate function to analyze the EE of
strongly interacting states, especially CFTs, on tori. This is a powerful tool since virtually no other analytic results
exist in this case. In an important development, a semi-analytical result was obtained\cite{fradkin} for
$\tor$ in special CFTs using the holographic AdS/CFT 
correspondence. However, singular behavior was found as the aspect ratio
goes through $b\!=\!1$. Such non-analycities are not expected for generic CFTs, as in the quantum critical Ising model,
and are indeed absent in \req{tor-emi} and in the free boson CFT (\rfig{scalar_2d}).   
Nevertheless, as noted above, striking similarities exist for $b\leq 1$ between the EMI and AdS functions, \rfig{comparison}.   
In the latter case $b\cdot(\tor-\tor(\pi))$ is exactly independent of $b$\cite{fradkin}, 
while for the EMI, this holds to excellent accuracy
and is not a priori obvious from \req{tor-emi}, see App.\ts\ref{ap:properties_EMI} for more details.    

A useful limit to consider is the thin torus: $\asp\!\to\!\infty$ with $\thA$ fixed,
in which case \req{tor-emi} reduces to $2\infcyl + O(b^{-2})$. Namely, \bl{the universal EE} approaches a pure
constant independent of $L_A,L_i$, \bl{which is twice the universal EE associated with a semi-infinite cylindrical bipartition of an infinite cylinder (App.~\ref{ap:EMI}).}
This is consistent with the expectation that a generic
CFT will not contain gapless modes in the 1d limit \bl{because the contracting $y$-direction leads to a large $\sim\! 1/L_y$ gap.}
Otherwise, \bl{the EE} would scale as $\sim\ln[\tfrac{L_x}{\pi\delta}\sin(\tfrac{\pi L_A}{L_x})]$ \bl{when $L_A$ is changed, corresponding to the 
behavior of a critical 1d system  on a circle}\cite{Calabrese1}.     
The \bl{absence of such critical scaling in the thin torus limit} is verified\cite{fradkin} in the strongly coupled holographic CFTs
mentioned above.  
Exceptions do occur, \eg for non-interacting CFTs with \emph{periodic} boundary conditions due to zero \bl{energy} modes, but
one can twist the boundary conditions to \bl{gap them out} (see below). 
  
\begin{figure}  
\center 
   \includegraphics[scale=0.39]{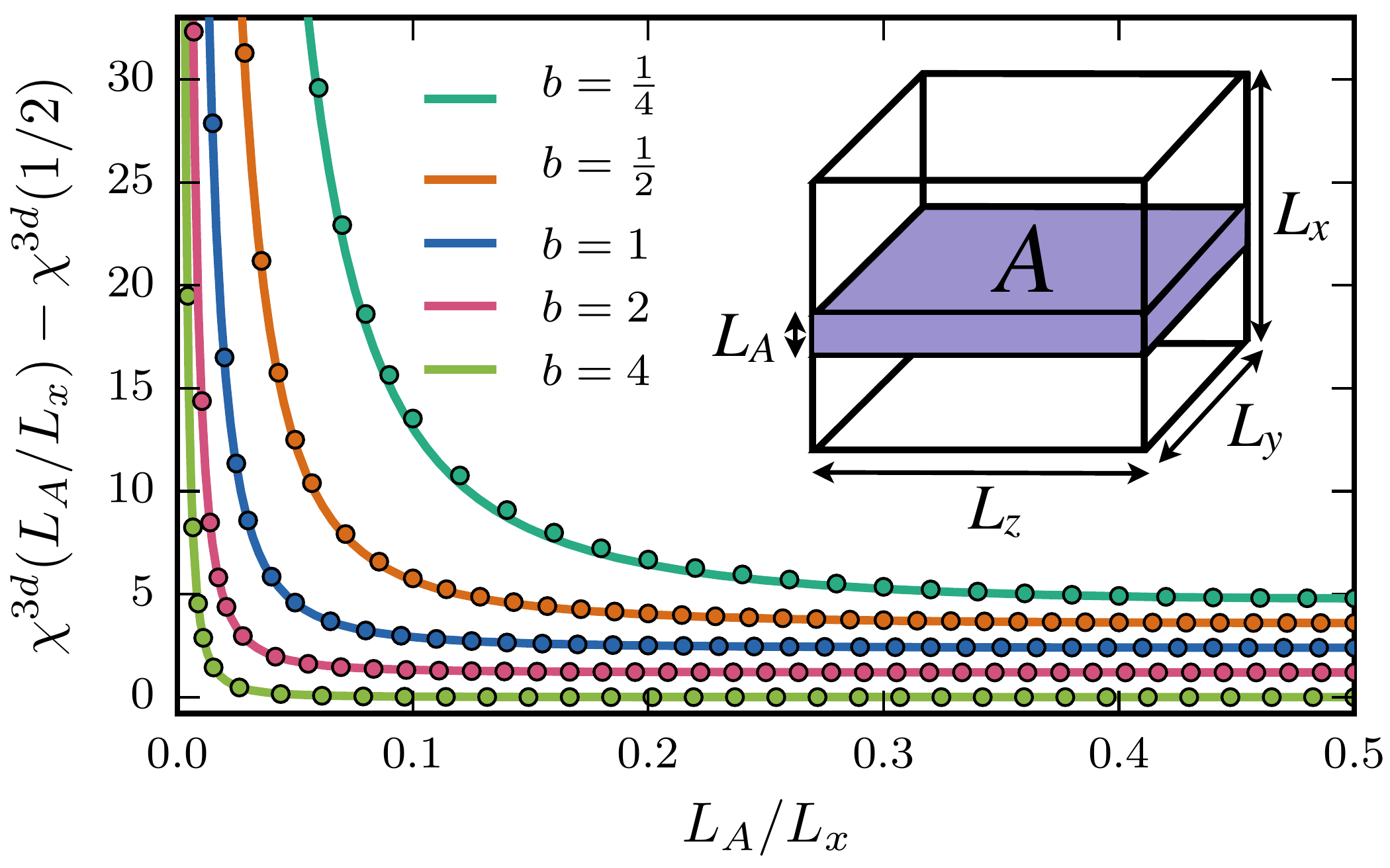} 
   \caption{{\bf Main}: Torus function $\tort$ for the massless free boson in 3d \bl{for various aspect ratios $b_y\!=\! b_z\!=\! b$}.  
The points are the numerical data 
and the line is the prediction obtained using the
ansatz $\tortEMI$, \emph{without any fitting parameters}. {\bf Inset}: Opposing faces
of the box are identified to give a 3-torus topology. 
Region $A$ is a hyper-cylinder extending along $x$.}
\label{fig:scalar_3d}
\centering         
\end{figure}

{\bf 3d torus:}  
We now explore the largely uncharted territory of torus entanglement in gapless 3d theories.
We take the subregion $A$ to be a hyper-cylinder of length $L_A$ aligned along $x$,
\rfig{scalar_3d}.
The corresponding angle variable is again $\thA=2\pi L_A/L_x$. The analog of \req{SA} in 3d reads:  
\begin{align} \label{SA3d}
  S^{3d}(A)= \mathcal B\frac{2 L_yL_z}{\delta^2}-\tort +O(\delta/L_{y,z}), 
\end{align}
where $\tort(\thA;\asp_y,\asp_z)$ now depends on the 2 aspect ratios, $b_{y,z}=L_x/L_{y,z}$. 
The general properties obtained above for the 2d torus function $\tor$
can be adapted \emph{mutatis mutandis} to the 3d case. In particular, $\tort$ will be
convex decreasing for $0\!<\!\thA\!\leq\! \pi$, as in \req{ineq}, and will be analytic about $\thA\!=\!\pi$. 
Further, in the small-$\thA$ limit 
we find
\begin{align} \label{small_3d}
  \tort(\thA\to 0) = \kappat \, \frac{L_y L_z}{L_A^2} = \frac{(2\pi)^2\kappat}{b_y b_z\, \theta^2} 
\end{align} 
since the EE effectively becomes that of an infinite thin slab with thickness $L_A$. 
Our 2d argument given above can be generalized to argue that the  
system is insensitive to the periodicity of the $x,y,z$ directions in this limit. 
$\kappat\geq 0$ is a universal constant characterizing the theory\cite{Casini_rev}, and it is the 3d analog of 
the 2d $\kappa$ encountered above.  
As we have done in 2d, we can use the EMI to obtain a closed-form torus function $\tortEMI(\thA)$.
\bl{\rfig{scalar_3d} shows the result for different aspect ratios}; the full answer is given in App.\ts\ref{ap:EMI}.    

{\bf Torus EE for lattice bosons:} 
We numerically evaluate the torus EE of a free and massless relativistic boson (a CFT) in 2d/3d
using the square/cubic lattice realization of the Hamiltonian
$H\! =\! \int \! d^d x  [ \tfrac{1}{2}\pi^2 + \tfrac{1}{2}( \nabla \phi)^2 ]$,
where $\phi$ is the 1-component boson and $\pi$ its conjugate momentum.
This theory corresponds to the Gaussian fixed point of the interacting quantum critical Ising model in 2d/3d, and constitutes a key benchmark system. We obtain the torus EE \bl{by directly evaluating the reduced density matrix of $A$} from the two-point  
vacuum correlation functions 
$\left\langle \phi_\mathbf{x} \phi_\mathbf{x'} \right\rangle$ and 
$\left\langle \pi_\mathbf{x} \pi_\mathbf{x'} \right\rangle$ 
for lattice sites $\mathbf{x},\mathbf{x'}\!\in\! A$ \cite{Peschel};
details are given in App.\ts\ref{ap:numerics}.
We perform our 2d calculations on lattices of size \bl{$L_x=500$, 
and our 3d calculations on lattices of size $L_x = 100, 140, 200, 288, 456$
for aspect ratios 
$b _y\!=\! b_z \!= \tfrac{1}{4}, \tfrac{1}{2},1,2,4$ (respectively).}
Each lattice has antiperiodic boundary conditions (APBC) in the $y$-direction and PBC along the remaining directions. 
We use the former to avoid the $\mathbf{k}\!=\!0$ zero mode present for PBC. 

The numerical results in 2d/3d are shown in Figs.\ts\ref{fig:scalar_2d},\ref{fig:scalar_3d}, respectively.  
The solid lines in both figures correspond to the EMI candidate functions, 
Eq.\ts(\ref{tor-emi}) in 2d, while the 3d one is given in the appendix. 
Crucially, \emph{no fitting} to the data has been performed. Instead, to generate the lines we have relied on two facts:
First, the EMI torus functions relative to their value at $\theta\!=\!\pi$,  $\torEMI^{3d}(\theta)-\torEMI^{3d}(\pi)$,  
depend on a single universal constant, $\kappa^{3d}$. Second, this constant has been 
computed in a \emph{different} context for the massless scalar in 2d/3d\cite{Casini_rev}: $\kappa_{\rm sc} = 0.0397$, 
$\kappat_{\rm sc} =5.54\! \times\! 10^{-3}$.  
The resulting ansatz curves and the data agree with each other exceptionally well, which
is surprising since we have not done any fitting. The agreement in \bl{2d/3d} extends over a wide 
range of aspect ratios, meaning the ansatz even captures the $b$-dependence \bl{without any fitting}! 
We note that since the EMI does not describe a free boson CFT, 
we expect that some of the deviations are intrinsic.    

The inset of \rfig{scalar_2d} shows the data for a massless free Dirac fermion (another CFT) obtained 
numerically in Ref.\ts\onlinecite{fradkin} with (A)PBC along $(y)x$.      
In this case, we again know the value of the small-$\thA$ constant\cite{Casini_rev}, $\kappa_{\rm Dirac}=0.0722$,
which allows us to fix the $\torEMI$ ansatz; the result is the line in the inset of \rfig{scalar_2d}.
We have also shown the data for a complex scalar, which overlaps almost exactly with that of the Dirac fermion.
Part of the agreement can be explained from the fact that the complex scalar has $\kappa\!=\!2\kappa_{\rm sc}\!=\!0.0794$,
which is close to the Dirac value.   

{\bf Summary \& outlook:}   
We have seen that the universal EE of cylindrical regions on tori reveals non-trivial
information about scale invariant quantum systems, like conformal field theories, in 2d/3d. 
Our findings range from general non-perturbative properties to concrete examples involving bosons on a lattice.
We note that many of these results can be extended to the
R\'enyi entropies $S_n$. In particular, in the thin slice limit $\tor_n$ will show the same divergence 
as in Eqs.\ts(\ref{thin},\ref{small_3d}), but with $\ka_n$. 
A torus function was previously derived\cite{Stephan13} at $n\!\geq \! 2$ for a family
of 2d Lifshitz quantum critical points\cite{Ardonne04}, and it was successfully compared with the von Neumann case 
in various theories. Many of our results apply to that function\cite{will-prep}.

\bl{Since the torus EE can also capture topological information about the excitations\cite{Dong08,Zhang12,Cincio} (relating to anyons, say),
it will be interesting to use it to obtain
fingerprints for gapless spin liquids or deconfined quantum critical points.} 
In this vein, a simple example where $\tor$ encodes both topological and geometrical degrees of freedom is 
Kitaev's gapless spin liquid on the honeycomb lattice\cite{Kitaev06}. In this frustrated spin model, the emergent long-distance degrees
of freedom are 2 massless Majorana fermions coupled to a $Z_2$ gauge field. 
We expect the universal EE to be $\tor_{\rm f}(\thA)+\tor_{\rm top}$,
owing to the factorization of the fermions and $Z_2$ contributions\cite{Yao10}.    
$\tor_{\rm top}$ is purely topological and comes from the $Z_2$ gauge theory\cite{Zhang12},  
while the fermions yield the shape dependent $\tor_{\rm f}(\thA)$.
Inspired by this capability of $\chi$ to capture both topological and gapless degrees of freedom,
we ask ask whether the torus EE can yield a RG monotone, in the same spirit as the disk EE\cite{sinha10,CH12}?       
 
\emph{Acknowledgments}---We are thankful to P.~Bueno, X.~Chen,
E.~Fradkin, A.~Lucas for useful discussions.   
WWK is grateful for the hospitality of Perimeter Institute, where this work was initiated.
WWK was funded by a fellowship from NSERC, and by MURI grant W911NF-14-1-0003 from ARO.  
LHS was partially funded by the Ontario Graduate Scholarship.
RM is supported by NSERC of Canada, the Canada Research Chair Program and 
the Perimeter Institute for Theoretical Physics. 
Research at Perimeter Institute is supported by the Government of Canada through the Department of Innovation, Science and Economic Development and by the Province of Ontario through the Ministry of Research \& Innovation. 
   
\appendix   

\begin{center} 
{\Large\bf Supplementary Information} 
\end{center} 

\tableofcontents        

\section{Torus entanglement of the Extensive Mutual Information model}
\label{ap:EMI}  

\begin{figure}
\center
   \includegraphics[scale=.38]{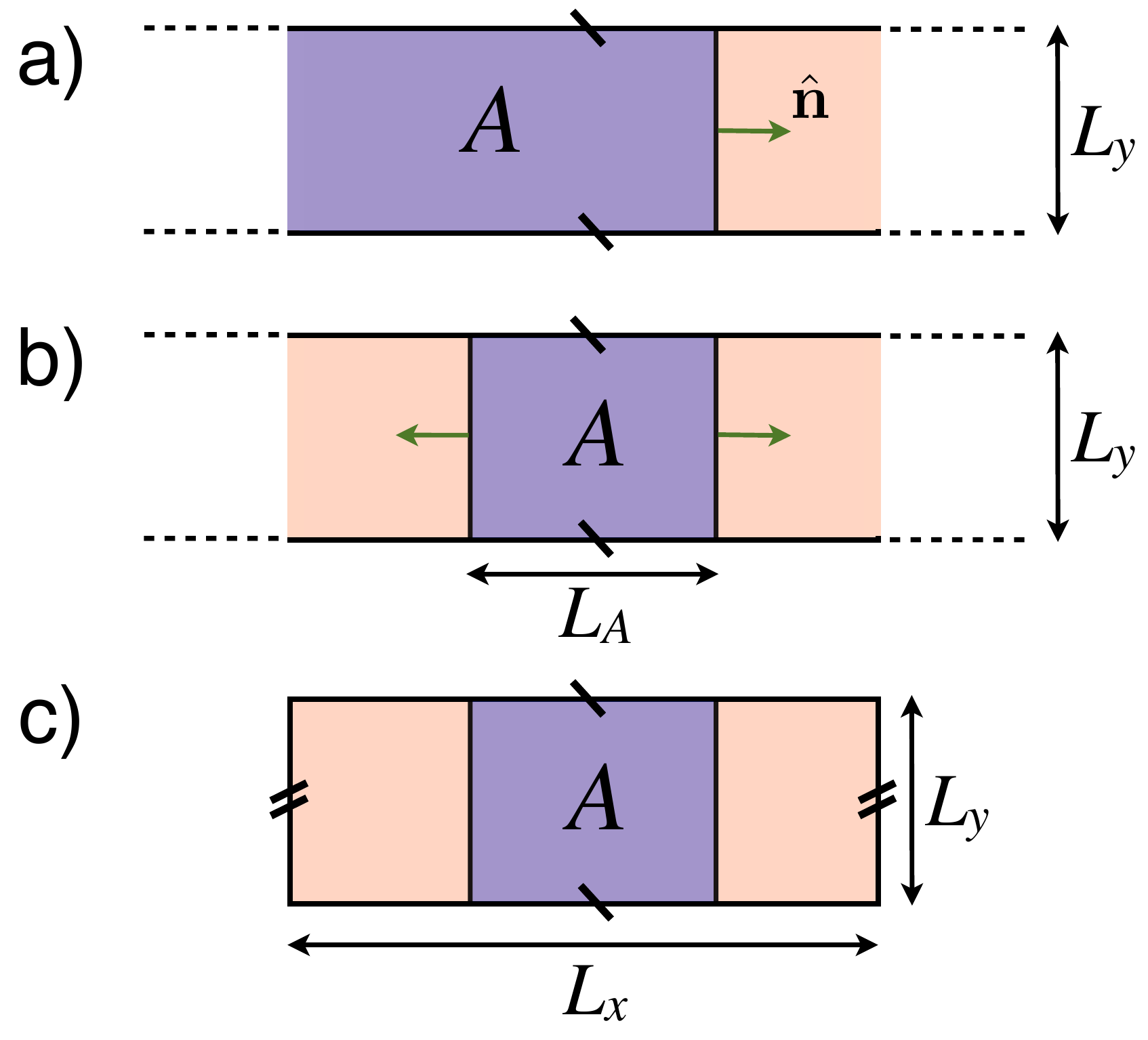}      
   \caption{a) Semi-infinite subregion $A$ of a space with the topology of an infinite cylinder.
b) Finite subregion $A$ of the infinite cylinder.
c) Finite subregion $A$ of a space with a torus topology.
Each green arrow denotes a normal vector, $\hat{\b n}$, to the entangling surface.
} 
\label{fig:cyl}
\centering         
\end{figure}    

Let us first recall the calculation of the EE within the Extensive Mutual Information
model (EMI) for a region $A$ in  
infinite flat space\cite{Casini05,Casini08,Swingle10}. 
One needs to evaluate the following double integral over two copies of the entangling surface $\partial A$:
\begin{align}  \label{s-emi}
    S(A) &= \int_{\partial A}\! d \b r_1\! \int_{\partial A} \! d \b r_2 \, \hat n_1\cdot \hat n_2 \, C(\b r_1-\b r_2), 
\end{align}
with
\begin{align}
  C(\b r) &=\frac{s_1}{|\b r|^{2(d-1)}},  \label{emi2}
\end{align}  
where $d$ is the spatial dimension, $\hat{\b n}$ the vector normal to $\partial A$, and $\b r_{12}=\b r_1-\b r_2$ is the separation vector.
\bl{The prescription given in \req{emi2} cannot be applied to the torus because it does not account for the periodicity of space. 
On the torus we require $C(\b r)$ to be periodic along all the spatial directions. Further, at distances much shorter
than the linear dimensions of the torus, $L_i$, the $C$-function needs to reduce to \req{emi2}, i.e.\ 
\begin{align}
  C(r_i\ll L_i ) = s_1/|\b r|^{2(d-1)}\,,
\end{align}
because in this limit the boundary conditions of space should not affect the EE. 
This scaling naturally follows from the scale invariance of the system on the infinite plane\cite{Casini05,Casini08,Swingle10}. 
We note that under such general conditions, $S(A)$ defined in \req{s-emi} will have an extensive mutual information $I$ for non-intersecting regions $A,B,C$
living on the torus: 
\begin{align}
  I(A,B\cup C) = I(A,B) + I(A,C)
\end{align}
where $I(A,B)=S(A)+S(B) - S(A\cup B)$. We are thus led to the conclusion that on the torus, the form of $C(\b r)$
is not entirely constrained by symmetry and extensivity of the mutual information. This is not very surprising because
conformal symmetry on the torus is much less powerful than in infinite space $\mathbb R^d$.
For instance, conformal symmetry in 1+1 Euclidean spacetime dimensions
is not sufficiently powerful to fix the 2-point functions of local operators in a CFT on the torus; they will thus depend on many details of each
given theory. In contrast, such correlation functions are fixed by symmetry on the infinite plane. 

We thus see that the structure of the EMI on the torus is richer than on the plane. In order to obtain
a viable EMI ansatz, we shall construct a $C$-function that extends \req{emi2} to the torus in a simple way. 
In Section~\ref{sec:new-emi}, we construct a second EMI ansatz using a different $C(\b r)$, which has very similar properties to the first one.
This fact, together with the excellent agreement with the numerical data for the boson and Dirac fermion on the lattice, show the robustness
and utility of the EMI construction.
}

\subsection{Infinite cylinder} 
Let us begin with the simpler case where the entire space takes the topology of an infinite cylinder with circumference $L_y$, \rfig{cyl}a.

\begin{center} 
\emph{a. Semi-infinite region $A$}
\end{center}

We first take the region $A$ to be a semi-infinite cylinder which ends at $x=0$. The entangling surface is a circle,
and constitutes the domain of integration in the EMI calculation.
The expression for the EE entropy reads 
\begin{align} 
  S &= \int_0^{L_y}\! dy_1 \int_0^{L_y}\! dy_2 \, \hat n_1\cdot \hat n_2 \, C(\b r_1-\b r_2) \\
    &= \int_0^{L_y}\! dy_1 \int_0^{L_y}\! dy_2 \, s_1 \frac{\hat x\cdot\hat x}{\tilde y_{12}^2}, \label{cyl1}
\end{align}
where we have defined $y_{12}=y_1-y_2$, and
\begin{align} \label{ytilde}  
  |\tilde y| = 
  \begin{cases} 
    |y|, & \mbox{if  }\, |y|\leq L_y/2 \\ 
    L_y - |y|, & \mbox{if  }\,  L_y/2 < |y|\leq L_y,
  \end{cases} 
\end{align} 
\req{ytilde} defines a rather simple distance function on a line segment with periodic boundary conditions, \ie a (``flat'') circle.  
This constitutes our minimal prescription in modifying \req{emi2} to account for the periodicity in the $y$-direction.  
\bl{We shall see that this ansatz yields sensible and transparent answers for the EE.}
In fact, our results for the EE on the cylinder and torus topologies will be shown to satisfy all the known requirements (see main text). Further, they will provide very accurate candidate functions to compare with numerical data for non-interacting bosons and fermions.   

In obtaining \req{cyl1}, we have used  
the fact that both normal vectors are $\hat{\b n}_{1,2}=\hat x$, and $x_{1,2}=0$ on the entangling surface.  
We change variables to $Y=(y_1+y_2)/2$ and $y_{12}$, and perform the integral over $Y$:
\begin{align}
  S &= s_1L_y \int_{-L_y}^{L_y} \frac{dy_{12}}{\tilde y_{12}^2}
       = s_1L_y\, 4\!\int_{\de}^{L_y/2} \frac{dy_{12}}{y_{12}^2},
\end{align}
where in the last equality we have used the definition of $\tilde y_{12}$, \req{ytilde}. 
We also introduced a UV cutoff $\de$ to make the integral finite. The final result for the EE
in the limit $L_y\gg \de$ reads 
\begin{align} \label{semi-inf}
  S = \mathcal B \frac{L_y}{\de} - \infcyl + \dotsb,
\end{align}
where $\mathcal B=4s_1$,  
and $\infcyl=8 s_1$. Both constants are positive since $s_1>0$. 
We have thus recovered the boundary law term, and a universal (negative) contribution $-\infcyl$. 

\begin{center} 
\emph{b. Finite region $A$}
\end{center}
Let us now consider the more interesting case where the subregion $A$ is a cylinder of finite
length $L_A$, as shown in \rfig{cyl}b. 
The boundary of region $A$ is now composed of 2 disjoint circles (left and right): $\partial A=L\cup R$.
The EE computed within the EMI will thus be composed of 4 contributions, depending on
whether $\b r_i$ lies on the left or right circle: 
\begin{align} \label{LR}
  S &= S^{LL}+ S^{RR} + S^{LR} + S^{RL} \nn
    &= 2S^{RR} + 2S^{RL}.
\end{align}
In the last equality we have used the translation symmetry along the $x$-direction. 
Now, $S^{RR}$ is exactly given by the semi-infinite cylinder answer, \req{semi-inf}.   
It remains to compute $S^{RL}$:
\begin{align}
  S^{RL} = \int_0^{L_y}\! dy_1 \int_0^{L_y}\! dy_2 \, \frac{-s_1}{L_A^2 + \tilde y_{12}^2}, 
\end{align}
where we have used $\hat{\b n}_1=-\hat{\b n}_2$ when
$\b r_{1,2}$ do not lie on the same circle (disconnected component of $\partial A$).  
By again changing variables to $Y$ and $y_{12}$,
and performing the integrals we obtain 
\begin{align}
  S^{RL} = - s_1\, \frac{4L_y}{L_A} \cot\inv\! \left( \frac{2L_A}{L_y} \right).
\end{align}
Note that this result is entirely independent of the cutoff $\de$. The final answer for the EE of $A$ thus reads:
\begin{align}
  S = \mathcal B \frac{2L_y}{\de} - \cyl + \dotsb ,
\end{align}
where we have defined the cylinder function $\cyl(L_A/L_y)$,  
\begin{align} \label{cyl-emi}
    \cyl &= 2\infcyl + \frac{2\kappa}{\pi} \frac{L_y}{L_A}\cot\inv\! \left( \frac{2L_A}{L_y} \right),   
\end{align}
which depends on the single dimensionless ratio, $L_A/L_y$. Here $\kappa=4\pi s_1$ 
is the ``thin strip'' constant: It determines the subleading term in the EE 
of a thin strip of width $L_A$ in the infinite plane\cite{Casini_rev}, 
\begin{align}
  S_{\rm strip} = \mathcal B \frac{2L}{\de} - \kappa \frac{L}{L_A} +\dotsb ,
\end{align}
where $L$ is a long-distance regulator for region $A$.
$\mathcal B$ and $\infcyl$ are as above, in particular $\infcyl=8s_1$.   
\begin{figure}
\center
   \includegraphics[scale=1]{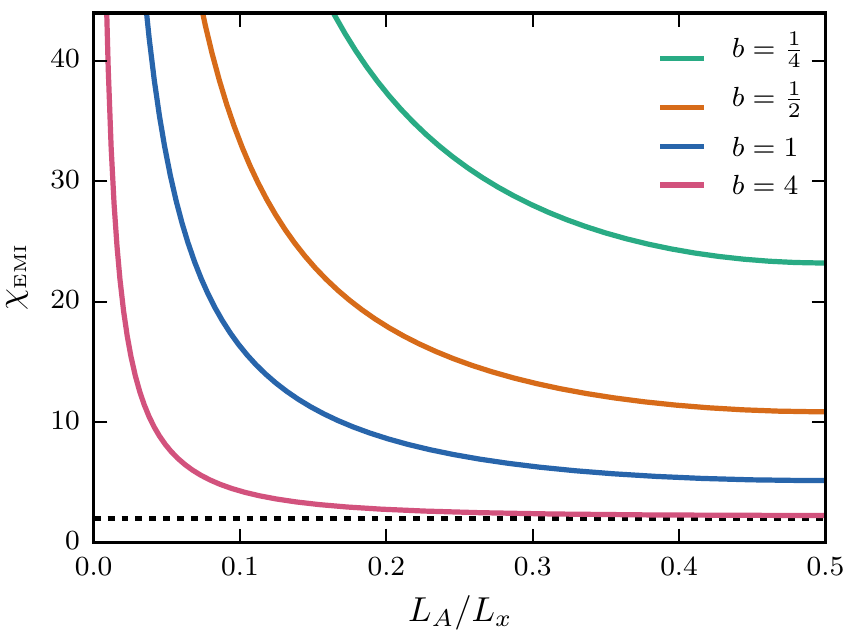} 
   \caption{Full $\torEMI$ (no vertical offset) 
for different aspect ratios (using $s_1=1/8$). From bottom to top: $L_x/L_y=4,1,1/2,1/4$. 
The horizontal dashed line corresponds to $L_x/L_y=\infty$, in which case $\torEMI=2\infcyl=2$.   
} 
\label{fig:EMI} 
\centering         
\end{figure}    
\subsection{Torus in 2d}   
We can now tackle the torus topology in 2d, \rfig{cyl}c. The extra complication compared with the infinite cylinder
is that the space is now periodic in the $x$-direction. As a result the final answer must be the same 
under the exchange $L_A\leftrightarrow L_x-L_A$ (by purity). The structure of the EE within in the EMI will
still decompose into 2 terms, as in \req{LR}. $S_{RR}$ will be the same as in the cylinder calculation
above because it is not sensitive to the $x$-cycle. In contrast, $S_{RL}$ does know about the $x$-cycle. 
The simplest way to accommodate for this is to add a ``mirror'' contribution to \req{emi2}, 
$C^{\rm mirror}=s_1 /[(L_x\!-\!L_A)^2+\tilde{y}_{12}^2]$, in the calculation of $S^{RL}=S^{LR}$. 
Performing the calculation with the additional mirror term, we finally obtain 
\begin{align}
  S = \mathcal B \frac{L_y}{\de} - \torEMI + \dotsb,
\end{align}
where the torus EE function is
\begin{align}
  \torEMI = 2\infcyl + \frac{2\kappa L_y}{\pi}\! \left[ \frac{\cot\inv\!\left( \frac{{2L_A}_{}}{L_y}\right) }{L_A}
 +  \frac{\cot\inv\!\left( \frac{2(L_x-L_A{)}_{} }{L_y}\right) }{L_x-L_A}  \right]\!
\end{align} 
and $\mathcal B=4 s_1$, $\infcyl=8s_1$ and $\kappa=4\pi s_1$ are the same constants as in the
infinite cylinder calculations above. 
$\mathcal B$ is independent of the aspect ratio of the torus. 
Further, $\torEMI$ is non-negative. \rfig{EMI} shows the $b,\thA$-dependence of $\torEMI$.

\subsection{Torus in 3d}
We now turn to the torus topology in 3d, see \rfig{scalar_3d}. 
We compactify the spatial dimensions such that the $i$th coordinate 
$r_i$ is identified with $r_i+L_i$, $i=x,y,z$.
The entangling surface consists of 2 disconnected parts, each of which is a 2-torus.
As a result, the EE within the EMI again decomposes into 2 terms as in \req{LR}.  
The first term, $S^{RR}$, comes from having $\b r_1$ and $\b r_2$ both on the right 2-torus $R$:
\begin{align}
  S^{RR} = \int dy_1 dz_1 \int dy_2 dz_2 \frac{s_1}{\left(\tilde y_{12}^2 + \tilde z_{12}^2\right)^2}, 
\end{align}
where $\tilde y_{12}$ is as defined in \req{ytilde}, $\tilde z_{12}$ is analogously defined but
with $L_z$ instead of $L_y$. We change integration variables to $Y=(y_1+y_2)/2$, $y_{12}$, and
$Z=(z_1+z_2)/2$, $z_{12}$, and perform the integrals over $Y,Z$:
\begin{align}
  S^{RR}
    = 16L_yL_z\! \int_{0}^{\frac{{L_y} }{2}}\!\! dy_{12}\int_{0}^{\frac{{L_z}_{} }{2}}\!\! dz_{12} 
  \frac{s_1}{\left(y_{12}^2 + z_{12}^2\right)^2},   
\end{align}
where we were able to remove the tildes by restricting the domain of integration. 
Performing both integrals we get
\begin{multline}
  S^{RR} = 2\pi s_1 \frac{L_yL_z}{\de^2} - 16s_1\left[ 1+\frac{L_y}{L_z}\tan\inv\!\left(\frac{L_y}{L_z}\right)\right. \\ \left.
    +\frac{L_z}{L_y}\tan\inv\!\left(\frac{L_z}{L_y}\right) \right] + O(\de/L_{y,z}),
\end{multline}
which is symmetric under $L_y\leftrightarrow L_z$, as expected. The term in brackets is independent of the
UV cutoff $\de$, and will contribute to the torus function $\tortEMI$.

Next, we turn to the $S^{RL}$ term in \req{LR}. As in the 2d torus calculation, $S^{RL}$   
will receive a contribution from a term with $|x_{12}|=L_A$, and from a ``mirror'' term with
$|x_{12}|=L_x-L_A$, in order to account for the periodicity along $x$. 
The first contribution reads 
\begin{align}
  S^{RL}_{(1)}= \int dy_1 dz_1 \int dy_2 dz_2 \frac{-s_1}{\left(L_A^2+ \tilde y_{12}^2 + \tilde z_{12}^2\right)^2} ,
\end{align}
where we have used $\hat{\b n}_1=-\hat{\b n}_2$. Again changing variables to center-of-mass and relative
coordinates, and performing the integration over the former we get
\begin{align}
   S^{RL}_{(1)} &= \! \int_0^{\frac{L_y}{2}} \!\! dy_{12}\int_0^{\frac{L_{z_{\!}}}{2}} \!\! dz_{12} 
  \frac{-16L_yL_zs_1}{\left(L_A^2 + y_{12}^2 + z_{12}^2\right)^2}   \\ 
    &=-\frac{16L_y L_z s_1}{2L_A^2}\! \left[ \frac{L_y \tan\inv\!\Big( \frac{ {L_z}_{} }{ \sqrt{4L_A^2+L_y^2} } \Big) }{\sqrt{4L_A^2+L_y^2}} + y\leftrightarrow z \right]\!.\nonumber
  \end{align}
Thus the final answer for the EE is 
\begin{align}
  S=\mathcal B\frac{2L_y L_z}{\de^2} - \tortEMI +\dotsb ,
\end{align}
where 
\begin{widetext}
  \begin{multline}  \label{torEMI-3d-full}
    \tortEMI =  2\infcyl^{3d} + \frac{4\kappat}{\pi}\!\left[ \frac{L_y}{L_z}\tan\inv\!\left(\frac{L_y}{L_z} \right) +
    \frac{L_y L_z}{2L_A^2}\frac{L_y \tan\inv\!\Big(\frac{{L_z}_{}}{\sqrt{4L_A^2+L_y^2}} \Big) }{\sqrt{4L_A^2+L_y^2}} \right. \\ 
\left. + \,  \frac{L_y L_z}{2 (L_x \!-\!L_A)^2 } \frac{L_y \tan\inv\!\Big(\frac{{L_z}_{}}{\sqrt{4(L_x-L_A)^2+L_y^2}} \Big) }{\sqrt{4(L_x\! -\! L_A)^2+L_y^2}} 
 + y\leftrightarrow z \right] 
  \end{multline}
\end{widetext}   
with $\infcyl^{3d}=16s_1$ and $\kappat=8\pi s_1$. We note that the square brackets contain
6 terms due to the contributions with $y$ and $z$ interchanged. 
Thus, $\tortEMI$ is fully symmetric under the exchange $L_y\leftrightarrow L_z$, as it should be. 
We have defined $\kappat$ as the thin slab coefficient:
\begin{align}
  \lim_{L_A\to 0} \torEMI = \kappat \frac{L_y L_z}{L_A^2},
\end{align}
which is readily obtained from \req{torEMI-3d-full} by using the identity 
$\tan\inv z +\tan\inv(1/z)=\pi/2$, with $z>0$.  
\bl{The result simplifies for $b_y=b_z=b$ to give
\begin{multline}  \label{tort-emi}
  \tortEMI(\thA) = 16\pi \kappat \Bigg[\frac{1}{8\pi} + \frac{\cot\inv\! \sqrt{1+(b\thA/\pi)^2}}{b^2\thA^2 \sqrt{1+(b\thA/\pi)^2}} \\
 +  \frac{\cot\inv\!\sqrt{1+b^2(2\pi-\thA)^2/\pi^2} }{b^2(2\pi-\thA)^2 \sqrt{1+b^2(2\pi-\thA)^2/\pi^2}} \Bigg] + 
2\infcyl^{3d}.
\end{multline}    
This is the case we study numerically in \rfig{scalar_3d} of the main text for the non-interacting gapless boson.
We note that $\tortEMI$ has a similar structure to the 2d result, \eg the appearance of $\cot\inv$.} 

\subsection{Properties in $2d$}
\label{ap:properties_EMI}

\subsubsection{Special scaling for $b \leq 1$}  
\begin{figure*}
\centering
 \subfigure[]{\label{fig:ReCtt} \includegraphics[scale=.9]{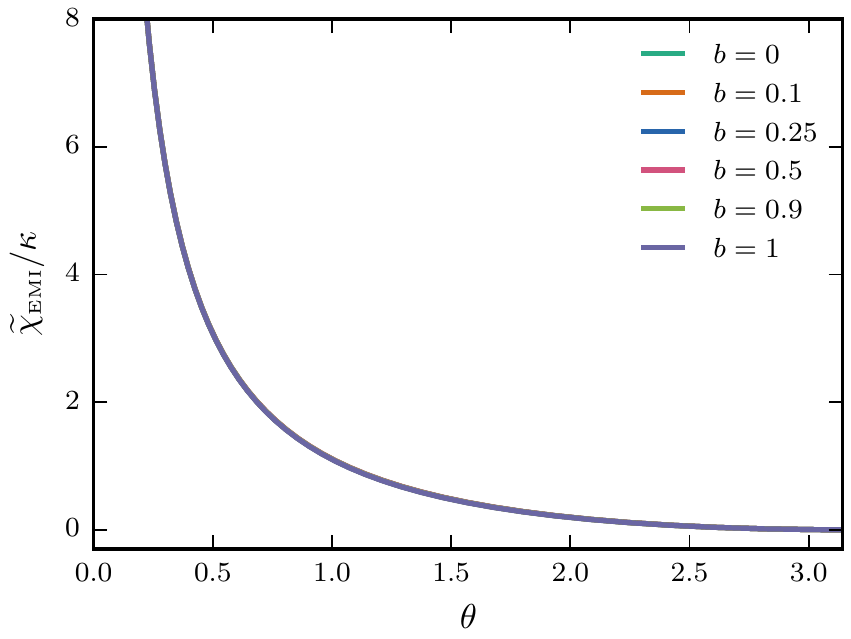}} 
 \subfigure[]{\label{fig:ImCtt} \includegraphics[scale=.9]{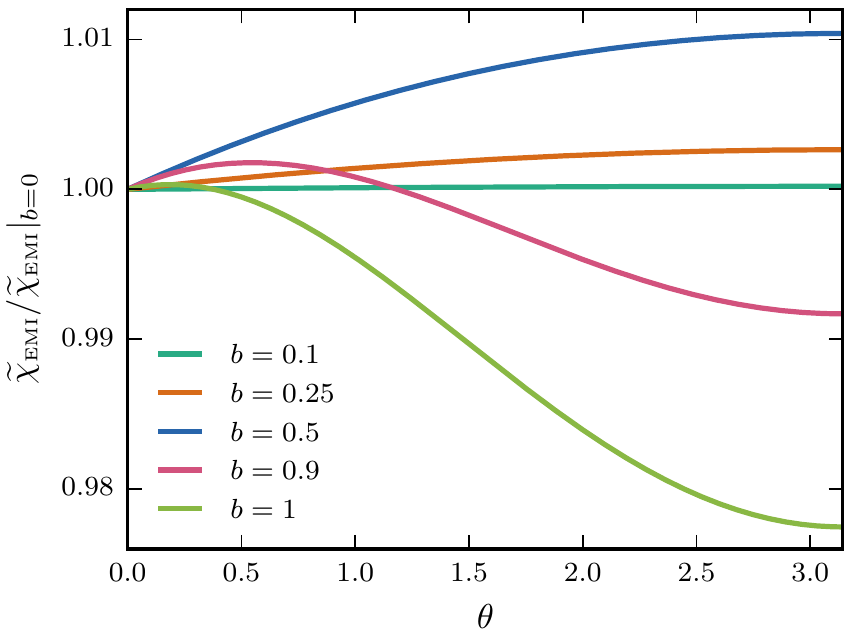}} 
\caption{\label{fig:chi-tilde}
{\bf a)} $\tEMIsp(\thA)/\kappa$ for different aspect ratios $b$. The curves are plotted in different colors
but cannot be distinguished in this plot. 
{\bf b)} For the same values of $b$
as in a), we plot $\tEMIsp$ divided by 
its value at $b=0$, $\tEMIsp(\thA;b)/\tEMIsp(\thA;0)$. Note the very narrow ordinate range!
} 
\end{figure*}

We discuss the special scaling encountered for $b\cdot\torEMI(\thA)$, relative to its value at $\thA=\pi$:
\begin{align}
  \tEMIsp(\theta) &=  \frac{b}{2\pi} \left[ \torEMI(\thA;b)-\torEMI(\pi;b) \right] \\
  &= \frac{4\kappa}{\pi} \left( \frac{\cot\inv(b\thA/\pi)}{\thA} + \frac{\cot\inv(b(2\pi- \thA)/\pi)}{2\pi-\thA} \right. \nn
  & \qquad \qquad \left. -\frac{2}{\pi}\cot\inv b\right) .
\end{align}
As was mentioned in the main text, $\tEMIsp$ is approximately independent of the aspect ratio $b$
in the range $b\leq 1$. \rfig{chi-tilde} demonstrates that this statement holds very accurately.  

We note that in the $b\to 0$ limit
\begin{align}
  \tEMIsp(\thA; 0) = \frac{2\kappa}{\pi} \, \frac{(\pi-\thA)^2}{\thA (2\pi-\thA)}.
\end{align}
We recognize this as precisely the corner EE function $a(\thA)$ of the family of Lifshitz quantum critical points 
with conformal wavefunctions\cite{Fradkin06}! For that family of $z=2$ theories\cite{Ardonne04}, 
$\kappa=\pi c/24$\cite{Fradkin06,WKB}, where 
$c$ is the Virasoro central charge of the parent 1d CFT that describes the equal-time correlations of the theory.     
At present, we do not have an explanation for the appearance of this corner function in the torus EE
of the EMI. We note that the latter has a different corner function $a_{\rm EMI}(\theta)$.\cite{Casini08,Swingle10}   

\subsubsection{Smooth and thin-torus limit expansions}   
We here give the leading terms in the smooth $\thA\approx\pi$ and thin slice $\thA\approx 0$ 
expansions for $\torEMI$ in 2d. In the smooth limit we get
\begin{multline}
  \torEMI(\thA\approx\pi)=
\left(2\infcyl + \frac{8 \kappa  \cot ^{-1}b}{\pi  b} \right) \\
+ \frac{8  \kappa  \left(b+2 b^3+\left(b^2+1\right)^2 \cot ^{-1}b\right)}{\pi ^3 b \left(b^2+1\right)^2}(\theta -\pi )^2 + O(\thA-\pi)^4
\end{multline}
while in the thin slice limit we get 
\begin{multline}
  \torEMI(\thA\approx 0)= \frac{2 \pi  \kappa }{b \theta } + \left(2\infcyl + \frac{2 \kappa  \cot ^{-1}(2 b)}{\pi  b} -\frac{4 \kappa }{\pi }\right) 
  +O(\thA).
\end{multline}
We see that the first term matches the thin strip contribution, as described above and in the main text.
\bl{
\subsection{Another ansatz}  \label{sec:new-emi}
The reader might wonder how generic is the ansatz described in the above sections? Indeed, one could have chosen another ansatz for the
$C(\b r)$ function in \req{s-emi}. Here, we introduce another ansatz, and show that it has very similar properties to the first one.
We focus on $d=2$, where the new ansatz for the $C$-function reads:
\begin{align} \label{new-emi}
  C(\b r) = \frac{s_1}{\chord(x)^2 + \chord(y)^2}
\end{align}
where 
\begin{align} \label{chord}
  \chord(r_i) = \frac{L_i}{\pi} \sin\left(\frac{\pi r_i}{L_i}\right) 
\end{align}
is the chord length on a circle of circumference $L_i$. \req{new-emi} is manifestly periodic  under $r_i\to r_i+L_i$, and
reduces to \req{emi2} at short distances. The resulting torus function $\chi$ can be easily obtained using \req{s-emi}: 
\begin{align}
  \chi(\theta;b) = \frac{\pi\kappa}{b\sin(\theta/2) \sqrt{1+b^2\sin^2(\theta/2)} }\,. 
\end{align}
This new ansatz, just like the original one, obeys all the known physical constraints required for $\chi$, such as being convex decreasing on $[0,\pi)$,
and the reflection property $\chi(2\pi-\theta)=\chi(\theta)$. Further, in \rfig{new-emi} we quantitatively compare the new ansatz with the old one,  
using $\chi(\theta)-\chi(\pi)$ normalized by $\kappa$. This is the same quantity that is compared with the boson and fermion numerical data
in the main text. We see that the deviations are very small.  
\begin{figure}
\center 
   \includegraphics[scale=.5]{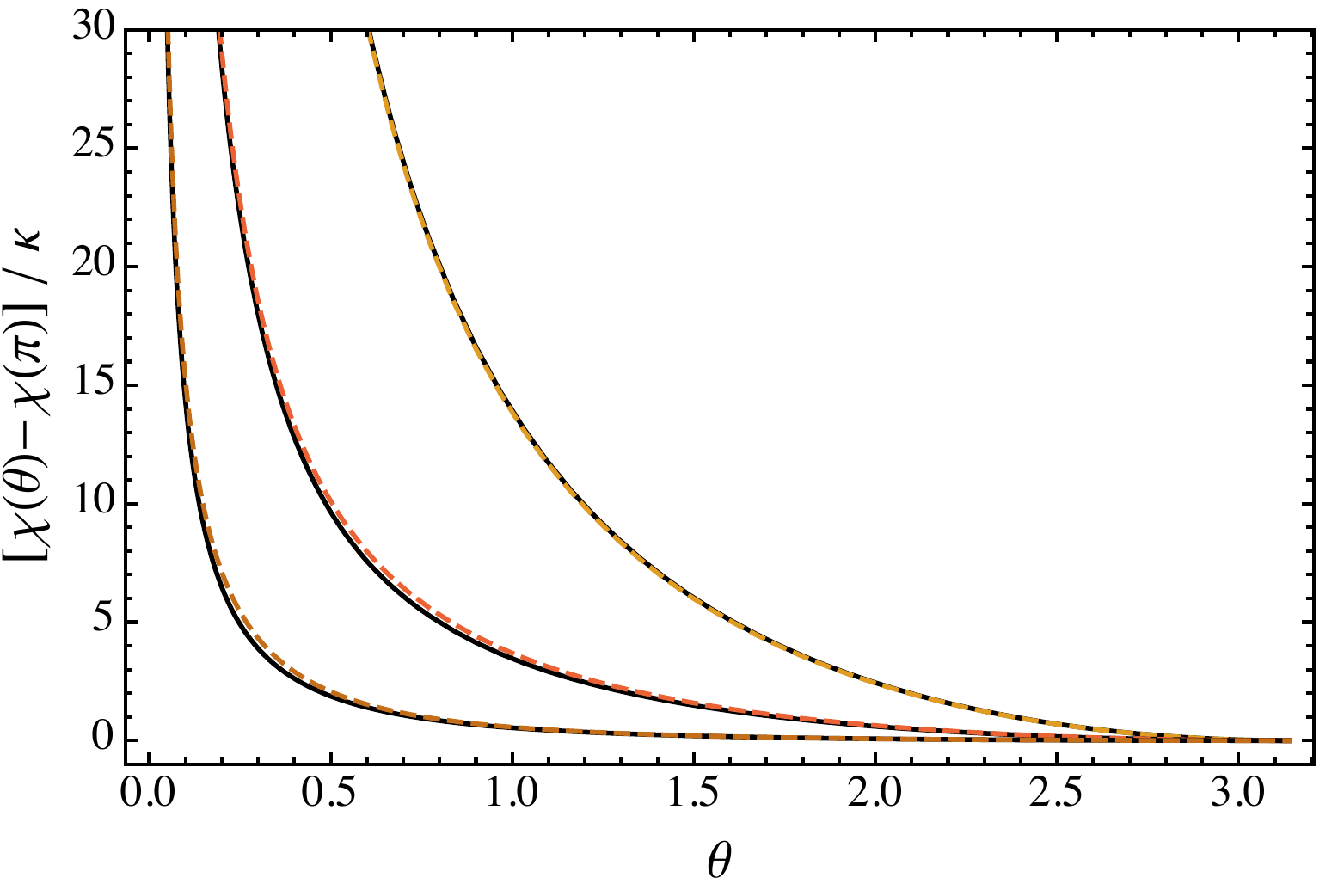}   
   \caption{We compare the new EMI ansatz (dashed) with the old one (black and solid) for 3 different aspect ratios. 
Top to bottom: $b=1/4,1,4$.}    
\label{fig:new-emi}
\centering         
\end{figure}        

}
\section{Numerical calculations} 
\label{ap:numerics}
We calculate the EE for the lattice Hamiltonian 
of a free relativistic boson in the massless limit, which is given by
\begin{align}
H = \frac{1}{2} \sum_\mathbf{x} 
& \Big[ \pi_{\mathbf{x}}^2 + \left( \phi_{x_1+1, x_2, \ldots, x_d} - \phi_{\mathbf{x}} \right)^2 \nonumber \\
\phantom{\bigg(} & {}  + \ldots + \left( \phi_{x_1, x_2, \ldots, x_d+1} - \phi_{\mathbf{x}} \right)^2 \Big],	\label{eq:Hamiltonian_lattice}
\end{align}
where $d$ is the spatial dimension of the lattice, 
$\mathbf{x} = (x_1, x_2, \ldots, x_d)$ represents the spatial lattice coordinates, 
each $x_i$ is summed from 1 to $L_i$,
and $L_i$ is the lattice length along the $i^\mathrm{th}$ dimension.

For translationally invariant boundary conditions, the two-point vacuum correlation functions corresponding to this Hamiltonian are given by
\begin{align}
\left\langle \phi_0 \phi_\mathbf{x} \right\rangle &= 
\frac{1}{2N} \sum_\mathbf{k}  
\frac{1}{\omega_\mathbf{k}} \cos \left( k_1 x_1 \right) \cos \left( k_2 x_2 \right) \cdots \cos \left( k_d x_d \right)
, \nonumber \\
\left\langle \pi_0 \pi_\mathbf{x} \right\rangle &= \frac{1}{2N} \sum_\mathbf{k} 
\omega_\mathbf{k} 
\cos \left( k_1 x_1 \right) \cos \left( k_2 x_2 \right) \cdots \cos \left( k_d x_d \right), \label{eq:correlators_PBC}
\end{align}
where $N = L_1 L_2 \cdots L_d$ is the total number of lattice sites and
\begin{equation}
\omega_\mathbf{k} = 2 \sqrt{ \sin^2\left( k_1/2 \right) +  \sin^2\left( k_2/2 \right)  + \ldots +  \sin^2\left( k_d/2 \right)}. \label{eq:omega}
\end{equation}
The values of the momenta $\mathbf{k}$ are quantized such that $k_i = 2 n_i \pi /L_i$ when the lattice has PBC along the $i^\mathrm{th}$ lattice direction and similarly $k_i = (2 n_i +1) \pi /L_i$ for APBC (where $n_i = 0, 1, \ldots L_i-1$).

Note that, for a fully periodic lattice, the correlator $\left\langle \phi_{0} \phi_\mathbf{x} \right\rangle$ (and, as we will see, the EE) diverges since $\omega_\mathbf{k}=0$ for the zero mode $\mathbf{k} = 0$.
In order to avoid this divergence, our calculations impose APBC along the $y$-direction (\ie the $x_2$-direction) and PBC along the remaining lattice directions. 
In doing so, we have $\omega_\mathbf{k} \neq 0$ for all allowed values of $\mathbf{k}$.

These two-point correlators define the \bl{$N \times N$} matrices 
$X_{ab} = \left\langle \phi_{\mathbf{x}_a} \phi_{\mathbf{x}_b} \right\rangle = 
\left\langle \phi_0 \phi_{\mathbf{x}_b-\mathbf{x}_a} \right\rangle$ 
and $P_{ab} = \left\langle \pi_0 \pi_{\mathbf{x}_b-\mathbf{x}_a} \right\rangle$, 
where $a,b$ label lattice sites. 
To get $S(A)$ we only need to know the \bl{smaller $N_A \times N_A$} matrices $X_A$ and $P_A$, 
which are the sections of $X$ and $P$ (respectively) with indices $i,j$ restricted to 
\bl{the $N_A$ sites of}
region $A$~\cite{Peschel}. 
The EE is given in terms of the eigenvalues $\nu_\ell$ of $\sqrt{X_A P_A}$ as\cite{Casini:2009}
\begin{align} 
S(A) = \sum_{\ell\bl{=1}}^{\bl{N_A}} & \left[ \left( \nu_\ell + \frac{1}{2} \right) \log \left( \nu_\ell + \frac{1}{2} \right) \right. \nonumber \\
& \;{} - \left. \left( \nu_\ell - \frac{1}{2} \right) \log \left( \nu_\ell - \frac{1}{2} \right) \right].
\end{align}

\bl{In order to access the EE on larger lattices, for the case of the torus geometry we employ an extension of the above methods
as given in Ref.~\onlinecite{fradkin} that takes advantage of the translational symmetry along $(d-1)$ spatial lattice directions. 
In this modified method, we map the $(d + 1)$-dimensional model to an effective model consisting of $L_2 \times L_3 \times ... \times L_d$ separate $(1+ 1)$-dimensional chains. }
 
\bibliographystyle{apsrev}
\bibliography{Biblo}

\begin{thebibliography}{54}
\expandafter\ifx\csname natexlab\endcsname\relax\def\natexlab#1{#1}\fi
\expandafter\ifx\csname bibnamefont\endcsname\relax
  \def\bibnamefont#1{#1}\fi
\expandafter\ifx\csname bibfnamefont\endcsname\relax
  \def\bibfnamefont#1{#1}\fi
\expandafter\ifx\csname citenamefont\endcsname\relax
  \def\citenamefont#1{#1}\fi
\expandafter\ifx\csname url\endcsname\relax
  \def\url#1{\texttt{#1}}\fi
\expandafter\ifx\csname urlprefix\endcsname\relax\def\urlprefix{URL }\fi
\providecommand{\bibinfo}[2]{#2}
\providecommand{\eprint}[2][]{\url{#2}}

\bibitem[{\citenamefont{Calabrese and Cardy}(2004)}]{Calabrese1}
\bibinfo{author}{\bibfnamefont{P.}~\bibnamefont{Calabrese}} \bibnamefont{and}
  \bibinfo{author}{\bibfnamefont{J.~L.} \bibnamefont{Cardy}},
  \bibinfo{journal}{J. Stat. Mech.} \textbf{\bibinfo{volume}{0406}},
  \bibinfo{pages}{P06002} (\bibinfo{year}{2004}), \eprint{hep-th/0405152}.

\bibitem[{\citenamefont{Casini and Huerta}(2009{\natexlab{a}})}]{Casini1}
\bibinfo{author}{\bibfnamefont{H.}~\bibnamefont{Casini}} \bibnamefont{and}
  \bibinfo{author}{\bibfnamefont{M.}~\bibnamefont{Huerta}},
  \bibinfo{journal}{J. Phys.} \textbf{\bibinfo{volume}{A42}},
  \bibinfo{pages}{504007} (\bibinfo{year}{2009}{\natexlab{a}}),
  \eprint{0905.2562}.

\bibitem[{\citenamefont{Fradkin}(2013)}]{Fradkin_book}
\bibinfo{author}{\bibfnamefont{E.}~\bibnamefont{Fradkin}},
  \emph{\bibinfo{title}{Field Theories of Condensed Matter Physics}}, Field
  Theories of Condensed Matter Physics (\bibinfo{publisher}{Cambridge
  University Press}, \bibinfo{year}{2013}), ISBN \bibinfo{isbn}{9780521764445}.

\bibitem[{\citenamefont{{Zeng} et~al.}(2015)\citenamefont{{Zeng}, {Chen},
  {Zhou}, and {Wen}}}]{Wen_book2}
\bibinfo{author}{\bibfnamefont{B.}~\bibnamefont{{Zeng}}},
  \bibinfo{author}{\bibfnamefont{X.}~\bibnamefont{{Chen}}},
  \bibinfo{author}{\bibfnamefont{D.-L.} \bibnamefont{{Zhou}}},
  \bibnamefont{and} \bibinfo{author}{\bibfnamefont{X.-G.} \bibnamefont{{Wen}}},
  \bibinfo{journal}{ArXiv e-prints}  (\bibinfo{year}{2015}),
  \eprint{1508.02595}.

\bibitem[{\citenamefont{{Laflorencie}}(2015)}]{laflorencie}
\bibinfo{author}{\bibfnamefont{N.}~\bibnamefont{{Laflorencie}}},
  \bibinfo{journal}{ArXiv e-prints}  (\bibinfo{year}{2015}),
  \eprint{1512.03388}.

\bibitem[{\citenamefont{{Dong} et~al.}(2008)\citenamefont{{Dong}, {Fradkin},
  {Leigh}, and {Nowling}}}]{Dong08}
\bibinfo{author}{\bibfnamefont{S.}~\bibnamefont{{Dong}}},
  \bibinfo{author}{\bibfnamefont{E.}~\bibnamefont{{Fradkin}}},
  \bibinfo{author}{\bibfnamefont{R.~G.} \bibnamefont{{Leigh}}},
  \bibnamefont{and}
  \bibinfo{author}{\bibfnamefont{S.}~\bibnamefont{{Nowling}}},
  \bibinfo{journal}{Journal of High Energy Physics}
  \textbf{\bibinfo{volume}{5}}, \bibinfo{eid}{016} (\bibinfo{year}{2008}),
  \eprint{0802.3231}.

\bibitem[{\citenamefont{{Zhang} et~al.}(2012)\citenamefont{{Zhang}, {Grover},
  {Turner}, {Oshikawa}, and {Vishwanath}}}]{Zhang12}
\bibinfo{author}{\bibfnamefont{Y.}~\bibnamefont{{Zhang}}},
  \bibinfo{author}{\bibfnamefont{T.}~\bibnamefont{{Grover}}},
  \bibinfo{author}{\bibfnamefont{A.}~\bibnamefont{{Turner}}},
  \bibinfo{author}{\bibfnamefont{M.}~\bibnamefont{{Oshikawa}}},
  \bibnamefont{and}
  \bibinfo{author}{\bibfnamefont{A.}~\bibnamefont{{Vishwanath}}},
  \bibinfo{journal}{\prb} \textbf{\bibinfo{volume}{85}}, \bibinfo{eid}{235151}
  (\bibinfo{year}{2012}), \eprint{1111.2342}.

\bibitem[{\citenamefont{Cincio and Vidal}(2013)}]{Cincio}
\bibinfo{author}{\bibfnamefont{L.}~\bibnamefont{Cincio}} \bibnamefont{and}
  \bibinfo{author}{\bibfnamefont{G.}~\bibnamefont{Vidal}},
  \bibinfo{journal}{Phys. Rev. Lett.} \textbf{\bibinfo{volume}{110}},
  \bibinfo{pages}{067208} (\bibinfo{year}{2013}).

\bibitem[{\citenamefont{{Isakov} et~al.}(2011)\citenamefont{{Isakov},
  {Hastings}, and {Melko}}}]{isakov}
\bibinfo{author}{\bibfnamefont{S.~V.} \bibnamefont{{Isakov}}},
  \bibinfo{author}{\bibfnamefont{M.~B.} \bibnamefont{{Hastings}}},
  \bibnamefont{and} \bibinfo{author}{\bibfnamefont{R.~G.}
  \bibnamefont{{Melko}}}, \bibinfo{journal}{Nature Physics}
  \textbf{\bibinfo{volume}{7}}, \bibinfo{pages}{772} (\bibinfo{year}{2011}),
  \eprint{1102.1721}.

\bibitem[{\citenamefont{{Wang} et~al.}(2011)\citenamefont{{Wang}, {Gu}, {Gong},
  and {Sheng}}}]{Wang11}
\bibinfo{author}{\bibfnamefont{Y.-F.} \bibnamefont{{Wang}}},
  \bibinfo{author}{\bibfnamefont{Z.-C.} \bibnamefont{{Gu}}},
  \bibinfo{author}{\bibfnamefont{C.-D.} \bibnamefont{{Gong}}},
  \bibnamefont{and} \bibinfo{author}{\bibfnamefont{D.~N.}
  \bibnamefont{{Sheng}}}, \bibinfo{journal}{Physical Review Letters}
  \textbf{\bibinfo{volume}{107}}, \bibinfo{eid}{146803} (\bibinfo{year}{2011}),
  \eprint{1103.1686}.

\bibitem[{\citenamefont{{Jiang} et~al.}(2012)\citenamefont{{Jiang}, {Wang}, and
  {Balents}}}]{jiang12}
\bibinfo{author}{\bibfnamefont{H.-C.} \bibnamefont{{Jiang}}},
  \bibinfo{author}{\bibfnamefont{Z.}~\bibnamefont{{Wang}}}, \bibnamefont{and}
  \bibinfo{author}{\bibfnamefont{L.}~\bibnamefont{{Balents}}},
  \bibinfo{journal}{Nature Physics} \textbf{\bibinfo{volume}{8}},
  \bibinfo{pages}{902} (\bibinfo{year}{2012}), \eprint{1205.4289}.

\bibitem[{\citenamefont{Depenbrock et~al.}(2012)\citenamefont{Depenbrock,
  McCulloch, and Schollw\"ock}}]{depenbrock}
\bibinfo{author}{\bibfnamefont{S.}~\bibnamefont{Depenbrock}},
  \bibinfo{author}{\bibfnamefont{I.~P.} \bibnamefont{McCulloch}},
  \bibnamefont{and}
  \bibinfo{author}{\bibfnamefont{U.}~\bibnamefont{Schollw\"ock}},
  \bibinfo{journal}{Phys. Rev. Lett.} \textbf{\bibinfo{volume}{109}},
  \bibinfo{pages}{067201} (\bibinfo{year}{2012}).

\bibitem[{\citenamefont{{Fradkin} and {Moore}}(2006)}]{Fradkin06}
\bibinfo{author}{\bibfnamefont{E.}~\bibnamefont{{Fradkin}}} \bibnamefont{and}
  \bibinfo{author}{\bibfnamefont{J.~E.} \bibnamefont{{Moore}}},
  \bibinfo{journal}{Physical Review Letters} \textbf{\bibinfo{volume}{97}},
  \bibinfo{eid}{050404} (\bibinfo{year}{2006}), \eprint{cond-mat/0605683}.

\bibitem[{\citenamefont{{St{\'e}phan} et~al.}(2009)\citenamefont{{St{\'e}phan},
  {Furukawa}, {Misguich}, and {Pasquier}}}]{Stephan09}
\bibinfo{author}{\bibfnamefont{J.-M.} \bibnamefont{{St{\'e}phan}}},
  \bibinfo{author}{\bibfnamefont{S.}~\bibnamefont{{Furukawa}}},
  \bibinfo{author}{\bibfnamefont{G.}~\bibnamefont{{Misguich}}},
  \bibnamefont{and}
  \bibinfo{author}{\bibfnamefont{V.}~\bibnamefont{{Pasquier}}},
  \bibinfo{journal}{\prb} \textbf{\bibinfo{volume}{80}}, \bibinfo{eid}{184421}
  (\bibinfo{year}{2009}), \eprint{0906.1153}.

\bibitem[{\citenamefont{{Hsu} et~al.}(2009)\citenamefont{{Hsu}, {Mulligan},
  {Fradkin}, and {Kim}}}]{Hsu09}
\bibinfo{author}{\bibfnamefont{B.}~\bibnamefont{{Hsu}}},
  \bibinfo{author}{\bibfnamefont{M.}~\bibnamefont{{Mulligan}}},
  \bibinfo{author}{\bibfnamefont{E.}~\bibnamefont{{Fradkin}}},
  \bibnamefont{and} \bibinfo{author}{\bibfnamefont{E.-A.} \bibnamefont{{Kim}}},
  \bibinfo{journal}{\prb} \textbf{\bibinfo{volume}{79}}, \bibinfo{eid}{115421}
  (\bibinfo{year}{2009}), \eprint{0812.0203}.

\bibitem[{\citenamefont{{Metlitski} et~al.}(2009)\citenamefont{{Metlitski},
  {Fuertes}, and {Sachdev}}}]{Max09}
\bibinfo{author}{\bibfnamefont{M.~A.} \bibnamefont{{Metlitski}}},
  \bibinfo{author}{\bibfnamefont{C.~A.} \bibnamefont{{Fuertes}}},
  \bibnamefont{and}
  \bibinfo{author}{\bibfnamefont{S.}~\bibnamefont{{Sachdev}}},
  \bibinfo{journal}{\prb} \textbf{\bibinfo{volume}{80}}, \bibinfo{eid}{115122}
  (\bibinfo{year}{2009}), \eprint{0904.4477}.

\bibitem[{\citenamefont{{Hsu} and {Fradkin}}(2010)}]{Hsu10}
\bibinfo{author}{\bibfnamefont{B.}~\bibnamefont{{Hsu}}} \bibnamefont{and}
  \bibinfo{author}{\bibfnamefont{E.}~\bibnamefont{{Fradkin}}},
  \bibinfo{journal}{Journal of Statistical Mechanics: Theory and Experiment}
  \textbf{\bibinfo{volume}{9}}, \bibinfo{pages}{09004} (\bibinfo{year}{2010}),
  \eprint{1006.1361}.

\bibitem[{\citenamefont{{Oshikawa}}(2010)}]{Oshikawa10}
\bibinfo{author}{\bibfnamefont{M.}~\bibnamefont{{Oshikawa}}},
  \bibinfo{journal}{ArXiv e-prints}  (\bibinfo{year}{2010}),
  \eprint{1007.3739}.

\bibitem[{\citenamefont{{Metlitski} and {Grover}}(2011)}]{Max11}
\bibinfo{author}{\bibfnamefont{M.~A.} \bibnamefont{{Metlitski}}}
  \bibnamefont{and} \bibinfo{author}{\bibfnamefont{T.}~\bibnamefont{{Grover}}},
  \bibinfo{journal}{ArXiv e-prints}  (\bibinfo{year}{2011}),
  \eprint{1112.5166}.

\bibitem[{\citenamefont{{Swingle} and {Senthil}}(2012)}]{Swingle12}
\bibinfo{author}{\bibfnamefont{B.}~\bibnamefont{{Swingle}}} \bibnamefont{and}
  \bibinfo{author}{\bibfnamefont{T.}~\bibnamefont{{Senthil}}},
  \bibinfo{journal}{\prb} \textbf{\bibinfo{volume}{86}}, \bibinfo{eid}{155131}
  (\bibinfo{year}{2012}), \eprint{1109.3185}.

\bibitem[{\citenamefont{{Chen} et~al.}(2015)\citenamefont{{Chen}, {Cho},
  {Faulkner}, and {Fradkin}}}]{fradkin}
\bibinfo{author}{\bibfnamefont{X.}~\bibnamefont{{Chen}}},
  \bibinfo{author}{\bibfnamefont{G.~Y.} \bibnamefont{{Cho}}},
  \bibinfo{author}{\bibfnamefont{T.}~\bibnamefont{{Faulkner}}},
  \bibnamefont{and}
  \bibinfo{author}{\bibfnamefont{E.}~\bibnamefont{{Fradkin}}},
  \bibinfo{journal}{Journal of Statistical Mechanics: Theory and Experiment}
  \textbf{\bibinfo{volume}{2}}, \bibinfo{eid}{02010} (\bibinfo{year}{2015}),
  \eprint{1412.3546}.

\bibitem[{\citenamefont{{Pretko} and {Senthil}}(2015)}]{Pretko}
\bibinfo{author}{\bibfnamefont{M.}~\bibnamefont{{Pretko}}} \bibnamefont{and}
  \bibinfo{author}{\bibfnamefont{T.}~\bibnamefont{{Senthil}}},
  \bibinfo{journal}{ArXiv e-prints}  (\bibinfo{year}{2015}),
  \eprint{1510.03863}.

\bibitem[{\citenamefont{Ju et~al.}(2012)\citenamefont{Ju, Kallin, Fendley,
  Hastings, and Melko}}]{Ju_2012}
\bibinfo{author}{\bibfnamefont{H.}~\bibnamefont{Ju}},
  \bibinfo{author}{\bibfnamefont{A.~B.} \bibnamefont{Kallin}},
  \bibinfo{author}{\bibfnamefont{P.}~\bibnamefont{Fendley}},
  \bibinfo{author}{\bibfnamefont{M.~B.} \bibnamefont{Hastings}},
  \bibnamefont{and} \bibinfo{author}{\bibfnamefont{R.~G.} \bibnamefont{Melko}},
  \bibinfo{journal}{Phys. Rev. B} \textbf{\bibinfo{volume}{85}},
  \bibinfo{pages}{165121} (\bibinfo{year}{2012}).

\bibitem[{\citenamefont{Inglis and Melko}(2013)}]{Inglis_2013}
\bibinfo{author}{\bibfnamefont{S.}~\bibnamefont{Inglis}} \bibnamefont{and}
  \bibinfo{author}{\bibfnamefont{R.~G.} \bibnamefont{Melko}},
  \bibinfo{journal}{New Journal of Physics} \textbf{\bibinfo{volume}{15}},
  \bibinfo{pages}{073048} (\bibinfo{year}{2013}).

\bibitem[{\citenamefont{Kulchytskyy et~al.}(2015)\citenamefont{Kulchytskyy,
  Herdman, Inglis, and Melko}}]{Bohdan}
\bibinfo{author}{\bibfnamefont{B.}~\bibnamefont{Kulchytskyy}},
  \bibinfo{author}{\bibfnamefont{C.~M.} \bibnamefont{Herdman}},
  \bibinfo{author}{\bibfnamefont{S.}~\bibnamefont{Inglis}}, \bibnamefont{and}
  \bibinfo{author}{\bibfnamefont{R.~G.} \bibnamefont{Melko}},
  \bibinfo{journal}{Phys. Rev. B} \textbf{\bibinfo{volume}{92}},
  \bibinfo{pages}{115146} (\bibinfo{year}{2015}).

\bibitem[{\citenamefont{Helmes and Wessel}(2014)}]{Helmes2014}
\bibinfo{author}{\bibfnamefont{J.}~\bibnamefont{Helmes}} \bibnamefont{and}
  \bibinfo{author}{\bibfnamefont{S.}~\bibnamefont{Wessel}},
  \bibinfo{journal}{Phys. Rev. B} \textbf{\bibinfo{volume}{89}},
  \bibinfo{pages}{245120} (\bibinfo{year}{2014}).

\bibitem[{\citenamefont{Luitz et~al.}(2015)\citenamefont{Luitz, Plat, Alet, and
  Laflorencie}}]{Luitz}
\bibinfo{author}{\bibfnamefont{D.~J.} \bibnamefont{Luitz}},
  \bibinfo{author}{\bibfnamefont{X.}~\bibnamefont{Plat}},
  \bibinfo{author}{\bibfnamefont{F.}~\bibnamefont{Alet}}, \bibnamefont{and}
  \bibinfo{author}{\bibfnamefont{N.}~\bibnamefont{Laflorencie}},
  \bibinfo{journal}{Phys. Rev. B} \textbf{\bibinfo{volume}{91}},
  \bibinfo{pages}{155145} (\bibinfo{year}{2015}).

\bibitem[{\citenamefont{Laflorencie et~al.}(2015)\citenamefont{Laflorencie,
  Luitz, and Alet}}]{Laflorencie2015}
\bibinfo{author}{\bibfnamefont{N.}~\bibnamefont{Laflorencie}},
  \bibinfo{author}{\bibfnamefont{D.~J.} \bibnamefont{Luitz}}, \bibnamefont{and}
  \bibinfo{author}{\bibfnamefont{F.}~\bibnamefont{Alet}},
  \bibinfo{journal}{Phys. Rev. B} \textbf{\bibinfo{volume}{92}},
  \bibinfo{pages}{115126} (\bibinfo{year}{2015}).

\bibitem[{\citenamefont{Lieb and Ruskai}(1973)}]{Lieb73}
\bibinfo{author}{\bibfnamefont{E.~H.} \bibnamefont{Lieb}} \bibnamefont{and}
  \bibinfo{author}{\bibfnamefont{M.~B.} \bibnamefont{Ruskai}},
  \bibinfo{journal}{Journal of Mathematical Physics}
  \textbf{\bibinfo{volume}{14}} (\bibinfo{year}{1973}).

\bibitem[{\citenamefont{Casini and Huerta}(2009{\natexlab{b}})}]{Casini_rev}
\bibinfo{author}{\bibfnamefont{H.}~\bibnamefont{Casini}} \bibnamefont{and}
  \bibinfo{author}{\bibfnamefont{M.}~\bibnamefont{Huerta}},
  \bibinfo{journal}{J. Phys.} \textbf{\bibinfo{volume}{A42}},
  \bibinfo{pages}{504007} (\bibinfo{year}{2009}{\natexlab{b}}),
  \eprint{0905.2562}.

\bibitem[{\citenamefont{Casini and Huerta}(2007)}]{Casini3}
\bibinfo{author}{\bibfnamefont{H.}~\bibnamefont{Casini}} \bibnamefont{and}
  \bibinfo{author}{\bibfnamefont{M.}~\bibnamefont{Huerta}},
  \bibinfo{journal}{Nucl. Phys.} \textbf{\bibinfo{volume}{B764}},
  \bibinfo{pages}{183} (\bibinfo{year}{2007}), \eprint{hep-th/0606256}.

\bibitem[{\citenamefont{Hirata and Takayanagi}(2007)}]{Hirata07}
\bibinfo{author}{\bibfnamefont{T.}~\bibnamefont{Hirata}} \bibnamefont{and}
  \bibinfo{author}{\bibfnamefont{T.}~\bibnamefont{Takayanagi}},
  \bibinfo{journal}{JHEP} \textbf{\bibinfo{volume}{02}}, \bibinfo{pages}{042}
  (\bibinfo{year}{2007}), \eprint{hep-th/0608213}.

\bibitem[{\citenamefont{{Kallin} et~al.}(2013)\citenamefont{{Kallin}, {Hyatt},
  {Singh}, and {Melko}}}]{Kallin13}
\bibinfo{author}{\bibfnamefont{A.~B.} \bibnamefont{{Kallin}}},
  \bibinfo{author}{\bibfnamefont{K.}~\bibnamefont{{Hyatt}}},
  \bibinfo{author}{\bibfnamefont{R.~R.~P.} \bibnamefont{{Singh}}},
  \bibnamefont{and} \bibinfo{author}{\bibfnamefont{R.~G.}
  \bibnamefont{{Melko}}}, \bibinfo{journal}{Physical Review Letters}
  \textbf{\bibinfo{volume}{110}}, \bibinfo{eid}{135702} (\bibinfo{year}{2013}),
  \eprint{1212.5269}.

\bibitem[{\citenamefont{Kallin et~al.}(2014)\citenamefont{Kallin, Stoudenmire,
  Fendley, Singh, and Melko}}]{Kallin14}
\bibinfo{author}{\bibfnamefont{A.~B.} \bibnamefont{Kallin}},
  \bibinfo{author}{\bibfnamefont{E.~M.} \bibnamefont{Stoudenmire}},
  \bibinfo{author}{\bibfnamefont{P.}~\bibnamefont{Fendley}},
  \bibinfo{author}{\bibfnamefont{R.~R.~P.} \bibnamefont{Singh}},
  \bibnamefont{and} \bibinfo{author}{\bibfnamefont{R.~G.} \bibnamefont{Melko}},
  \bibinfo{journal}{Journal of Statistical Mechanics: Theory and Experiment}
  \textbf{\bibinfo{volume}{2014}}, \bibinfo{pages}{P06009}
  (\bibinfo{year}{2014}).

\bibitem[{\citenamefont{{Bueno}
  et~al.}(2015{\natexlab{a}})\citenamefont{{Bueno}, {Myers}, and
  {Witczak-Krempa}}}]{corner-prl}
\bibinfo{author}{\bibfnamefont{P.}~\bibnamefont{{Bueno}}},
  \bibinfo{author}{\bibfnamefont{R.~C.} \bibnamefont{{Myers}}},
  \bibnamefont{and}
  \bibinfo{author}{\bibfnamefont{W.}~\bibnamefont{{Witczak-Krempa}}},
  \bibinfo{journal}{Physical Review Letters} \textbf{\bibinfo{volume}{115}},
  \bibinfo{eid}{021602} (\bibinfo{year}{2015}{\natexlab{a}}),
  \eprint{1505.04804}.

\bibitem[{\citenamefont{{Bueno}
  et~al.}(2015{\natexlab{b}})\citenamefont{{Bueno}, {Myers}, and
  {Witczak-Krempa}}}]{corner-twist}
\bibinfo{author}{\bibfnamefont{P.}~\bibnamefont{{Bueno}}},
  \bibinfo{author}{\bibfnamefont{R.~C.} \bibnamefont{{Myers}}},
  \bibnamefont{and}
  \bibinfo{author}{\bibfnamefont{W.}~\bibnamefont{{Witczak-Krempa}}},
  \bibinfo{journal}{Journal of High Energy Physics}
  \textbf{\bibinfo{volume}{9}}, \bibinfo{eid}{91}
  (\bibinfo{year}{2015}{\natexlab{b}}), \eprint{1507.06997}.

\bibitem[{\citenamefont{Stoudenmire et~al.}(2014)\citenamefont{Stoudenmire,
  Gustainis, Johal, Wessel, and Melko}}]{Gustainis}
\bibinfo{author}{\bibfnamefont{E.~M.} \bibnamefont{Stoudenmire}},
  \bibinfo{author}{\bibfnamefont{P.}~\bibnamefont{Gustainis}},
  \bibinfo{author}{\bibfnamefont{R.}~\bibnamefont{Johal}},
  \bibinfo{author}{\bibfnamefont{S.}~\bibnamefont{Wessel}}, \bibnamefont{and}
  \bibinfo{author}{\bibfnamefont{R.~G.} \bibnamefont{Melko}},
  \bibinfo{journal}{Phys. Rev.} \textbf{\bibinfo{volume}{B90}},
  \bibinfo{pages}{235106} (\bibinfo{year}{2014}).

\bibitem[{\citenamefont{{Helmes} and {Wessel}}(2015)}]{Helmes2}
\bibinfo{author}{\bibfnamefont{J.}~\bibnamefont{{Helmes}}} \bibnamefont{and}
  \bibinfo{author}{\bibfnamefont{S.}~\bibnamefont{{Wessel}}},
  \bibinfo{journal}{\prb} \textbf{\bibinfo{volume}{92}}, \bibinfo{eid}{125120}
  (\bibinfo{year}{2015}), \eprint{1411.7773}.

\bibitem[{\citenamefont{Helmes et~al.}(2016)\citenamefont{Helmes,
  Hayward~Sierens, Chandran, Witczak-Krempa, and Melko}}]{free-corners}
\bibinfo{author}{\bibfnamefont{J.}~\bibnamefont{Helmes}},
  \bibinfo{author}{\bibfnamefont{L.~E.} \bibnamefont{Hayward~Sierens}},
  \bibinfo{author}{\bibfnamefont{A.}~\bibnamefont{Chandran}},
  \bibinfo{author}{\bibfnamefont{W.}~\bibnamefont{Witczak-Krempa}},
  \bibnamefont{and} \bibinfo{author}{\bibfnamefont{R.~G.} \bibnamefont{Melko}},
  \bibinfo{journal}{Phys. Rev. B} \textbf{\bibinfo{volume}{94}},
  \bibinfo{pages}{125142} (\bibinfo{year}{2016}).

\bibitem[{\citenamefont{Bueno and Myers}(2015)}]{Bueno2}
\bibinfo{author}{\bibfnamefont{P.}~\bibnamefont{Bueno}} \bibnamefont{and}
  \bibinfo{author}{\bibfnamefont{R.~C.} \bibnamefont{Myers}},
  \bibinfo{journal}{JHEP} \textbf{\bibinfo{volume}{08}}, \bibinfo{pages}{068}
  (\bibinfo{year}{2015}), \eprint{1505.07842}.

\bibitem[{\citenamefont{{Faulkner} et~al.}(2016)\citenamefont{{Faulkner},
  {Leigh}, and {Parrikar}}}]{faulkner15}
\bibinfo{author}{\bibfnamefont{T.}~\bibnamefont{{Faulkner}}},
  \bibinfo{author}{\bibfnamefont{R.~G.} \bibnamefont{{Leigh}}},
  \bibnamefont{and}
  \bibinfo{author}{\bibfnamefont{O.}~\bibnamefont{{Parrikar}}},
  \bibinfo{journal}{Journal of High Energy Physics}
  \textbf{\bibinfo{volume}{4}}, \bibinfo{eid}{88} (\bibinfo{year}{2016}),
  \eprint{1511.05179}.

\bibitem[{\citenamefont{Casini et~al.}(2005)\citenamefont{Casini, Fosco, and
  Huerta}}]{Casini05}
\bibinfo{author}{\bibfnamefont{H.}~\bibnamefont{Casini}},
  \bibinfo{author}{\bibfnamefont{C.~D.} \bibnamefont{Fosco}}, \bibnamefont{and}
  \bibinfo{author}{\bibfnamefont{M.}~\bibnamefont{Huerta}},
  \bibinfo{journal}{J. Stat. Mech.} \textbf{\bibinfo{volume}{0507}},
  \bibinfo{pages}{P07007} (\bibinfo{year}{2005}), \eprint{cond-mat/0505563}.

\bibitem[{\citenamefont{Casini and Huerta}(2009{\natexlab{c}})}]{Casini08}
\bibinfo{author}{\bibfnamefont{H.}~\bibnamefont{Casini}} \bibnamefont{and}
  \bibinfo{author}{\bibfnamefont{M.}~\bibnamefont{Huerta}},
  \bibinfo{journal}{JHEP} \textbf{\bibinfo{volume}{03}}, \bibinfo{pages}{048}
  (\bibinfo{year}{2009}{\natexlab{c}}), \eprint{0812.1773}.

\bibitem[{\citenamefont{Swingle}(2010)}]{Swingle10}
\bibinfo{author}{\bibfnamefont{B.}~\bibnamefont{Swingle}}
  (\bibinfo{year}{2010}), \eprint{1010.4038}.

\bibitem[{\citenamefont{Peschel}(2003)}]{Peschel}
\bibinfo{author}{\bibfnamefont{I.}~\bibnamefont{Peschel}}, \bibinfo{journal}{J.
  Phys. A: Math. Gen.} \textbf{\bibinfo{volume}{36}}, \bibinfo{pages}{L205}
  (\bibinfo{year}{2003}).

\bibitem[{\citenamefont{{St{\'e}phan} et~al.}(2013)\citenamefont{{St{\'e}phan},
  {Ju}, {Fendley}, and {Melko}}}]{Stephan13}
\bibinfo{author}{\bibfnamefont{J.-M.} \bibnamefont{{St{\'e}phan}}},
  \bibinfo{author}{\bibfnamefont{H.}~\bibnamefont{{Ju}}},
  \bibinfo{author}{\bibfnamefont{P.}~\bibnamefont{{Fendley}}},
  \bibnamefont{and} \bibinfo{author}{\bibfnamefont{R.~G.}
  \bibnamefont{{Melko}}}, \bibinfo{journal}{New Journal of Physics}
  \textbf{\bibinfo{volume}{15}}, \bibinfo{eid}{015004} (\bibinfo{year}{2013}),
  \eprint{1207.3820}.

\bibitem[{\citenamefont{{Ardonne} et~al.}(2004)\citenamefont{{Ardonne},
  {Fendley}, and {Fradkin}}}]{Ardonne04}
\bibinfo{author}{\bibfnamefont{E.}~\bibnamefont{{Ardonne}}},
  \bibinfo{author}{\bibfnamefont{P.}~\bibnamefont{{Fendley}}},
  \bibnamefont{and}
  \bibinfo{author}{\bibfnamefont{E.}~\bibnamefont{{Fradkin}}},
  \bibinfo{journal}{Annals of Physics} \textbf{\bibinfo{volume}{310}},
  \bibinfo{pages}{493} (\bibinfo{year}{2004}), \eprint{cond-mat/0311466}.

\bibitem[{\citenamefont{{Witczak-Krempa \emph{et. al.}}}(in
  preparation)}]{will-prep}
\bibinfo{author}{\bibfnamefont{W.}~\bibnamefont{{Witczak-Krempa \emph{et.
  al.}}}} (\bibinfo{year}{in preparation}).

\bibitem[{\citenamefont{{Kitaev}}(2006)}]{Kitaev06}
\bibinfo{author}{\bibfnamefont{A.}~\bibnamefont{{Kitaev}}},
  \bibinfo{journal}{Annals of Physics} \textbf{\bibinfo{volume}{321}},
  \bibinfo{pages}{2} (\bibinfo{year}{2006}), \eprint{cond-mat/0506438}.

\bibitem[{\citenamefont{{Yao} and {Qi}}(2010)}]{Yao10}
\bibinfo{author}{\bibfnamefont{H.}~\bibnamefont{{Yao}}} \bibnamefont{and}
  \bibinfo{author}{\bibfnamefont{X.-L.} \bibnamefont{{Qi}}},
  \bibinfo{journal}{Physical Review Letters} \textbf{\bibinfo{volume}{105}},
  \bibinfo{eid}{080501} (\bibinfo{year}{2010}), \eprint{1001.1165}.

\bibitem[{\citenamefont{{Myers} and {Sinha}}(2010)}]{sinha10}
\bibinfo{author}{\bibfnamefont{R.~C.} \bibnamefont{{Myers}}} \bibnamefont{and}
  \bibinfo{author}{\bibfnamefont{A.}~\bibnamefont{{Sinha}}},
  \bibinfo{journal}{\prd} \textbf{\bibinfo{volume}{82}}, \bibinfo{eid}{046006}
  (\bibinfo{year}{2010}), \eprint{1006.1263}.

\bibitem[{\citenamefont{{Casini} and {Huerta}}(2012)}]{CH12}
\bibinfo{author}{\bibfnamefont{H.}~\bibnamefont{{Casini}}} \bibnamefont{and}
  \bibinfo{author}{\bibfnamefont{M.}~\bibnamefont{{Huerta}}},
  \bibinfo{journal}{\prd} \textbf{\bibinfo{volume}{85}}, \bibinfo{eid}{125016}
  (\bibinfo{year}{2012}), \eprint{1202.5650}.

\bibitem[{\citenamefont{{Bueno} and {Witczak-Krempa}}(2016)}]{WKB}
\bibinfo{author}{\bibfnamefont{P.}~\bibnamefont{{Bueno}}} \bibnamefont{and}
  \bibinfo{author}{\bibfnamefont{W.}~\bibnamefont{{Witczak-Krempa}}},
  \bibinfo{journal}{\prb} \textbf{\bibinfo{volume}{93}}, \bibinfo{eid}{045131}
  (\bibinfo{year}{2016}), \eprint{1511.04077}.

\bibitem[{\citenamefont{Casini and Huerta}(2009{\natexlab{d}})}]{Casini:2009}
\bibinfo{author}{\bibfnamefont{H.}~\bibnamefont{Casini}} \bibnamefont{and}
  \bibinfo{author}{\bibfnamefont{M.}~\bibnamefont{Huerta}},
  \bibinfo{journal}{Journal of Physics A: Mathematical and Theoretical}
  \textbf{\bibinfo{volume}{42}}, \bibinfo{pages}{504007}
  (\bibinfo{year}{2009}{\natexlab{d}}).

\end{thebibliography}
 
\end{document}